\documentclass[12pt]{article}

\usepackage{amssymb,amsmath,graphicx,color}
\makeatletter\@addtoreset{equation}{section}
\makeatother

\begin{document}
\begin{flushright}
KIAS-P10034
\end{flushright}

\vspace{9mm}

\begin{center}
{{{\Large \bf Interaction between M2-branes and Bulk Form Fields}
}\\[17mm]
Yoonbai Kim$^{1}$,~~O-Kab Kwon$^{1}$,~~Hiroaki Nakajima$^{3,4}$,
~~D.~D. Tolla$^{1,2}$ \\[3mm]
{\it $^{1}$Department of Physics,~BK21 Physics Research Division,
~Institute of Basic Science,\\
$^{2}$University College,\\
Sungkyunkwan University, Suwon 440-746, Korea},\\[2mm]
{\it $^{3}$School of Physics, Korea Institute for Advanced Study,
Seoul 130-722, Korea},\\[2mm]
{\it $^{4}$Department of Physics, Kyungpook National University\\
Taegu, 702-701, Korea}
}\\[2mm]
{\tt yoonbai@skku.edu,~okab@skku.edu,~nakajima@kias.re.kr,~ddtolla@skku.edu}
\end{center}

\vspace{20mm}

\begin{abstract}
We construct the interaction terms between the world-volume fields of
multiple M2-branes and the 3- and 6-form fields in the context of
ABJM theory with U($N$)$\times$U($N$) gauge symmetry.
A consistency check is made in the simplest case of a single M2-brane
{\it i.e.}, our construction matches the known effective action of M2-brane
coupled to antisymmetric 3-form field.
We show that when dimensionally reduced, our couplings coincide with the effective
action of D2-branes coupled to R-R 3- and 5-form fields in type IIA string theory.
We also comment on the relation between a coupling with a specific 6-form
field configuration and the supersymmetry preserving
mass deformation in ABJM theory.
\end{abstract}

\newpage

\tableofcontents

\section{Introduction}\label{sec1}

Recently, the Lagrangian descriptions of
multiple M2-branes were found in the low energy limit,
which are the Bagger-Lambert-Gustavsson(BLG)
theory \cite{Bagger:2006sk, Gustavsson:2007vu} and the
Aharony-Bergman-Jafferis-Maldacena(ABJM) theory \cite{Aharony:2008ug}.
The BLG theory, which is equivalent to the ABJM theory with SU(2)$\times$SU(2)
gauge group~\cite{VanRaamsdonk:2008ft}, has $\mathcal{N}=8$ supersymmetry and
it is an effective theory of two M2-branes.
The ABJM theory with U($N$)$\times$U($N$) gauge group
has $\mathcal{N}=6$ supersymmetry and it describes the dynamics of $N$ parallel
M2-branes sitting at the singularity of a space with ${\mathbb Z}_k$ orbifold,
where $k$ appears as the Chern-Simons level in the theory.

Low energy dynamics of D-branes is also depicted by supersymmetric gauge theories,
{\it i.e.}, it is the Dirac-Born-Infeld(DBI) action
or the super Yang-Mills theory to the leading order in $\alpha'$-expansion.
In addition, D-branes can couple to the bulk supergravity fields.
The bosonic bulk fields include R-R form fields, which couple to the D-branes
through Wess-Zumino(WZ)-type action~\cite{Li:1995pq,Douglas:1995bn,Myers:1999ps}.
In the case of single D$p$-brane, the WZ-type coupling is only to
the R-R form fields of rank $p+1$ or less.
For multiple D$p$-branes the action can include the couplings to
all kinds of R-R form fields~\cite{Myers:1999ps}.

Analogous to the WZ-type coupling of D-branes in string theory, in M-theory,
the M2- and M5-branes couple to the 3-form and dual 6-form fields
in 11-dimensional supergravity.
According to the recent development of the world-volume theories
for multiple M2-branes, the coupling of M2-branes to
the bulk form fields were discussed in the context of both
the BLG theory~\cite{Li:2008eza,Ganjali:2009kt,Kim:2009nc}
and the ABJM theory~\cite{Lambert:2009qw,Sasaki:2009ij}.
For arbitrary $N$ stacked M2-branes the analyses have been made
in the restricted context: Ref.~\cite{Lambert:2009qw} dealt with
the coupling with constant form fields which survive in the infinite
tension limit of M2-branes and Ref.~\cite{Sasaki:2009ij} considered
only the case of single M2-brane.
Along this line it is timely to study the coupling of $N$
M2-branes to the bulk form fields which have arbitrary dependence
on the transverse scalar fields.

In our previous paper~\cite{Kim:2009nc}, we discussed the coupling between
M2-branes and the bulk form fields in BLG theory with the aforementioned
general setting.
After the compactification procedure of Mukhi-Papageorgakis (MP Higgsing
procedure)~\cite{Mukhi:2008ux,Ezhuthachan:2009sr},
we verified that our proposal reproduces the R-R form field couplings
to D2-branes as well as the linearized DBI action in type IIA string theory.
In this paper we naturally generalize our previous construction to multiple
M2-branes of arbitrary $N$, described by the
ABJM theory with U($N$)$\times$U($N$) gauge symmetry.
Guided by gauge invariance, we propose a general form of WZ-type coupling.
In our proposal the form fields can carry gauge indices since we assume
that the form fields depend on the transverse scalar fields in
(anti)bifundamental representations of the gauge group.
Since M-theory provides no direct guiding principle for the WZ-type coupling,
there seems no compelling reason to restrict the form fields to be
antisymmetric in the global indices. Reflecting consistency with the known
WZ-type couplings of type IIA string theory after the MP Higgsing, one can
constrain the M-theory form fields to appropriate forms.
To be specific, we impose some appropriate constraints to the 3- and 6-form
fields in M-theory
and obtain the relations among these form fields and the R-R and NS-NS
form fields in type IIA string theory.
These constraints lead to the antisymmetric property of the form fields
and the symmetrized matrix products in the resulting WZ-type
couplings~\cite{Myers:1999ps} of type IIA string theory.
A byproduct is the generation of the linearized DBI action of multiple D2-branes,
where the gauge field strength $\tilde{F}_{\mu\nu}$ and the
NS-NS 2-form field $\tilde{B}_{\mu\nu}$ appear in the gauge-covariant combination,
$\tilde{F}_{\mu\nu}+\frac{1}{2\pi\alpha'}\tilde{B}_{\mu\nu}$.

The MP Higgsing for the WZ-type coupling of the 3-form fields is completely
carried out, however the corresponding procedure for the case of the 6-form fields
is performed only up to the quadratic order terms in Planck length.
For the cubic order, there are too many possible terms
which appear in the MP Higgsing procedure and this makes
the derivation of explicit relations between the R-R 5-form
fields and the 6-form fields complicated. We do not carry out the reduction
procedure for this order terms, rather we guess the expected couplings from the results
of the lower order terms in Planck length.

This paper is organized as follows. In section 2, we propose the general
structure of the coupling between multiple M2-branes and the bulk form
fields.
In section 3, we compare the proposal with the well known coupling
for the case of a single M2-brane.
In section 4, we check our proposal by taking the reduction
to type IIA string theory, via the MP Higgs mechanism.
In section 5, we establish the relation between our WZ-type coupling and
the supersymmetry preserving mass deformation in ABJM theory.
Section 6 is devoted to conclusion and discussion.

\section{Couplings between M2-branes and Form Fields}\label{sec2}

Low energy effective world-volume theory of the stacked $N$ parallel
flat M2-branes is described by the ABJM action. Its bosonic sector
contains two gauge fields, $A_\mu$ and $\hat A_\mu$, of
U($N$)$\times$U($N$) gauge symmetry and four complex scalar fields $Y^A$,
$(A=1,2,3,4)$. The real and imaginary components of $Y^A$ represent
the eight transverse directions $X^{I},~(I=1,2,...,8)$,
\begin{align}\label{YA}
Y^A=X^A+iX^{A+4}.
\end{align}
The scalar fields are in the bifundamental
representation of the U($N$)$\times$U($N$) gauge symmetry while the gauge fields
$A_\mu$ and $\hat A_\mu$ are in the adjoint representations of the left U($N$) and the right
U($N$) gauge symmetries, respectively.
Including the gauge indices we can write the bosonic fields as
\begin{align}
(A_{\mu})^{a}_{\; b},\quad
({\hat A}_{\mu})^{{\hat a}}_{\;{\hat b}},\quad
(Y^{A})^{a}_{\; {\hat a}}
%=(X^{A}+iX^{A+4})^{a}_{\; {\hat a}}
\quad (Y^{\dagger}_{A})^{{\hat a}}_{\; a}.
\label{aayy}
\end{align}

Low energy effective description of M-theory is known to
be the 11-dimensional supergravity for which the bosonic fields are graviton $g_{MN}$,
3-form $C_{(3)}$, and 6-form $C_{(6)}$ fields. Those form fields can interact with
the world-volume fields through pull-back.
The corresponding interactions in 10-dimensional type II string theories
are described by WZ-type couplings~\cite{Li:1995pq,Douglas:1995bn,Myers:1999ps}.
However, in the scheme of M-theory, the form of such interaction is not yet known
except in BLG theory with SU(2)$\times$SU(2)
gauge symmetry~\cite{Li:2008eza,Kim:2009nc} and ABJM theory with constant
form fields~\cite{Lambert:2009qw}.
In this section we naturally extend the results in \cite{Kim:2009nc}
to the case of ABJM theory with U($N$)$\times$U($N$) gauge symmetry.

We start with a general gauge-invariant couplings of 3-form fields to
multiple M2-branes. In analogy with the WZ-type couplings in type II string theories
we restrict the couplings to terms which are linear in the form fields.
Our proposal for the 3-form fields action is
\begin{align}
S_{C}^{(3)} = \mu_2\int d^3x\, &\frac{1}{3!}
\epsilon^{\mu\nu\rho}\left\{{\rm Tr}\right\}\Big[ C_{\mu\nu\rho} +
3\lambda C_{\mu\nu A} D_\rho Y^A + 3\lambda^2 \big(C_{\mu AB} D_\nu
Y^A D_\rho Y^B + C_{\mu A\bar B} D_\nu Y^A D_\rho Y_B^\dagger \big)
\nonumber \\
&+ \lambda^3 (C_{ABC} D_\mu Y^A D_\nu Y^B D_\rho Y^C + C_{AB\bar C}
D_\mu Y^A D_\nu Y^B D_\rho Y_C^\dagger\big) + ({\rm c.c.}) \Big],
\label{act3}
\end{align}
where $\mu_{2}$ is M2-brane tension, $\lambda=2\pi  l_{{\rm
P}}^{3/2}$ with Planck length $l_{{\rm P}}$, and  $\{{\rm Tr}\}$
represents all possible contractions of gauge indices among the
3-form fields and transverse scalars, which give single
traces.\footnote{In our convention the unbarred global indices $A,
B, C,\, \cdots$ of the form fields are contracted with those of
bifundamental scalar fields while the barred global indices $\bar
A,\bar B,\bar C,\,\cdots$ are contracted with those of
antibifundamental scalar fields. The unbarred indices are always
located to the left of the barred ones.} The covariant derivative is
\begin{equation}
D_{\mu}Y^A = \partial_{\mu} Y^A + iA_{\mu}Y^A-iY^A{\hat A}_{\mu}.
\label{cvd}
\end{equation}
In each term the dependence of the pull-back of the 3-form field on
$Y^A$ and $Y_A^\dagger$ should be chosen in such a way that the term
contains equal number of bifundamental and antibifundamental fields
so that the action keeps manifest gauge invariance. The gauge
invariance is achieved if we assume that the pull-back of the form
fields can have multiple non-Abelian gauge indices which are allowed
only through their functional dependence on the transverse scalars.
This assumption is motivated by the analogy with R-R form field
couplings to multiple D$p$-branes in string
theory~\cite{Myers:1999ps}, where the pull-back of the R-R fields
are defined by just replacing the transverse coordinates with the
transverse scalar fields. In the latter case, the dependence of the
transverse scalars in the the pull-back of the R-R fields was
obtained by performing the non-Abelian Taylor
expansion~\cite{Myers:1999ps}.
For instance for a rank $n$ R-R form field it can be
written as
\begin{align}
P(\tilde{C}_{(n)})(\tilde{X})=
\tilde{C}_{(n)}(0)+\tilde{X}^{i}\partial_{i}\tilde{C}_{(n)}(0)
+\frac{1}{2!}\tilde{X}^{i}\tilde{X}^{j}\partial_{i}\partial_{j}
\tilde{C}_{(n)}(0)+\cdots,
\end{align}
where the product among $\tilde{X}^{i}$'s is a matrix product. The
expansion contains uncontracted gauge indices and,
in the coupling between the D$p$-branes and the R-R fields, the
gauge trace is taken including the pull-backed R-R form fields.
This procedure was confirmed to leading order by examining string
scattering amplitudes~\cite{Garousi:1998fg}.

In order to clarify the meaning of $\{{\rm Tr}\}$ we present some
terms in the action (\ref{act3}) including the gauge indices
explicitly,
\begin{align}\label{modtr}
\{{{\rm Tr}}\}\big(C_{\mu\nu A}D_\rho Y^A\big) &= (C_{\mu\nu A})^{\hat a}_a
(D_\rho Y^A)^a_{~\hat a},
\nonumber \\
\{{{\rm Tr}}\} \big(C_{\mu AB} D_\nu Y^A D_\rho Y^B\big) &= (C_{\mu
AB})^{\hat a\hat b}_{ab} (D_\nu Y^A)^{a}_{~\hat a} (D_\rho
Y^B)^{b}_{~\hat b} ,
\nonumber \\
\{{{\rm Tr}}\}\big(C_{\mu A\bar B} D_\nu Y^A D_\rho Y_B^\dagger
\big) &= (C_{\mu A\bar B})^{\hat a b}_{a\hat b} (D_\nu
Y^A)^a_{~\hat a} (D_\rho Y_B^\dagger)^{\hat b}_{~b},
\nonumber \\
\{{\rm Tr}\}\big(C_{AB\bar C} D_\mu Y^A D_\nu Y^B D_\rho
Y_C^\dagger\big) &=(C_{AB\bar C})^{\hat a\hat b c}_{ab\hat c} (D_\mu
Y^A)^a_{~\hat a} (D_\nu Y^B)^b_{~\hat b} (D_\rho Y_C^\dagger)^{\hat
c}_{~c},
\nonumber\\
\{{{\rm Tr}}\}\big(C_{ABC} D_\mu Y^A D_\nu Y^B D_\rho Y^C\big) &=
(C_{ABC})^{\hat a\hat b\hat c}_{abc} (D_\mu Y^A)^a_{~\hat a} (D_\nu
Y^B)^b_{~\hat b} (D_\rho Y^C)^c_{~\hat c}.
\end{align}
In order to understand this gauge index structure let us take into account
the Taylor expansion for the pull-backed form fields
in the context of M-theory. Due to the gauge invariance and the single traceness
of the corresponding coupling, possible terms in the Taylor expansion
for the form fields are restricted.
As an example, we expand $C_{\mu A\bar B}$ in terms of the transverse scalar fields,
\begin{align}
(C_{\mu A\bar B})_{a\hat b}^{\hat a b} =& C_{\mu A\bar B}(0)
\delta_{\hat b}^{\hat a}\delta_a^b +\partial_{{\bar C}}\partial_{D}C_{\mu
A\bar{B}}(0)(Y_C^\dagger Y^D)^{\hat a}_{~\hat b} \delta_a^b +
\partial_{C}\partial_{{\bar D}}\hat C_{\mu A\bar{B}}(0) \delta^{\hat a}_{\hat b}
(Y^C Y_D^\dagger )^{b}_{~a} + \cdots,
\label{B2}
\end{align}
where the form fields and their derivatives
are functions of the worldvolume coordinates but do not depend on
the transverse scalar fields.
After being pulled-backed to the worldvolume, uncontracted gauge indices appear
in the right-hand side of \eqref{B2} due to their dependence
on the transverse scalar fields.
This clearly verifies that Taylor expansion of the form fields generates
non-Abelian structure even though the bulk form fields themselves do not
carry non-Abelian gauge indices.
The same procedure can be applied
for the other pull-backed 3- and 6-form fields in order to understand the non-Abelian
gauge index structure.

From now on let us turn to the WZ-type coupling between M2-branes and 6-form fields.
Along the same line with the 3-form coupling, imposing the gauge invariance
and allowing only single trace couplings, the most general 6-form field
coupling which is  linear in the 6-form fields is given by
\begin{align}\label{act6}
S_{C}^{(6)} = &\mu_2'\int d^3x\, \frac{1}{3!}
\epsilon^{\mu\nu\rho} \left\{{\rm Tr}\right\}\Big( C_{\mu\nu\rho A
B\bar C} \beta^{AB}_{~C}
+3\lambda \big(C_{\mu\nu ABC\bar D} D_\rho Y^A \beta^{BC}_{~D}
+ C_{\mu\nu AB\bar C\bar D} D_\rho Y_C^\dagger \beta^{AB}_{~D}
\big)
\nonumber \\
&+3\lambda^2 \big(C_{\mu ABCD\bar E} D_\nu Y^A D_\rho Y^B \beta^{CD}_{~E}
+C_{\mu ABC\bar D\bar E} D_\nu Y^A D_\rho Y_D^\dagger \beta^{BC}_{~E}
+C_{\mu AB\bar C\bar D\bar E} D_\nu Y_C^\dagger D_\rho
Y_D^\dagger \beta^{AB}_{~E} \big)
\nonumber \\
&+\lambda^3\big(C_{ABCDE\bar F}D_\mu Y^A D_\nu Y^B D_\rho Y^C \beta^{DE}_{~F}
+C_{ABCD\bar E\bar F}D_\mu Y^A D_\nu Y^B D_\rho Y_E^\dagger
\beta^{CD}_{~F}
\nonumber \\
&+C_{ABC\bar D\bar E\bar F}D_\mu Y^A D_\nu Y_D^\dagger D_\rho
Y_E^\dagger \beta^{BC}_{~F}
+C_{AB\bar C\bar D\bar E\bar F}D_\mu Y_C^\dagger D_\nu
Y_D^\dagger D_\rho Y_E^\dagger \beta^{AB}_{~F}\big) + ({\rm
c}.{\rm c}.) \Big),
\end{align}
where $\mu_2'=\tau\lambda\mu_2$, $\tau$ is a dimensionless parameter
which will be fixed after reduction to type IIA string theory, and
$\beta^{AB}_{~C}\equiv\frac12(Y^AY_C^\dagger Y^B - Y^BY_C^\dagger Y^A)$.
We assumed that apart from their appearance in the covariant derivatives,
the transverse complex scalar fields explicitly enter the action only in terms of
this specific cubic product $\beta^{AB}_{~C}$ and its complex conjugate.
This assumption is motivated by the fact that $\beta^{AB}_{~C}$ and
its complex conjugate appear in the sextic potential of ABJM theory.
Actually $\beta^{AB}_{~C}$ corresponds to three commutator in hermitian
3-algebra formulation of the ABJM theory~\cite{Bagger:2008se}.

For every term in the action (\ref{act6}) the contraction of gauge indices
is performed as was done for the 3-form field coupling in (\ref{modtr}).
For instance,
\begin{align}\label{sixtr}
C_{AB\bar C\bar D\bar E\bar F}D_\mu &Y_C^\dagger D_\nu Y_D^\dagger
D_\rho Y_E^\dagger \beta^{AB}_{~F}\nonumber\\
&=(C_{AB\bar C\bar D\bar E\bar F})_{\hat a\hat b\hat c d}^{abc\hat
d}(D_\mu Y_C^\dagger)^{\hat a}_{a} (D_\nu Y_D^\dagger)^{\hat b}_{b}
(D_\rho Y_E^\dagger)^{\hat c}_{c} (\beta^{AB}_{~F})^d_{\hat d}.
\end{align}
Furthermore we should restrict the form field as in \eqref{B2}
in order to obtain single trace couplings.

It is not easy to show that the actions in \eqref{act3} and
\eqref{act6} are invariant under gauge transformation of the 3- and
6-form fields $C_{(n)}\to C_{(n)}+d\Lambda_{(n-1)}$ due to the
non-Abelian structure of the coupling. As was proven in the case of
R-R fields in string theory~\cite{Ciocarlie:2001qv, Adam:2003uq},
this issue of gauge invariance will be addressed through further
study~\cite{KKNT}.

\section{Consistency Check for Single M2-brane}\label{sec3}

In this section we will test our proposal in the previous section by comparing
with the well known effective action of a single M2-brane
in the presence of 3-form field ${\hat C}_{mnp}$.
The effective action is given by~\cite{Bergshoeff:1987cm}
\begin{align}
S_{11}=-\mu_{2}\int d^{3}\sigma\sqrt{-{\rm det}(\partial_{\mu}x^{m}
\partial_{\nu}x^{n}g_{mn})}+\frac{\mu_{2}}{3!}\int d^{3}\sigma
\epsilon^{\mu\nu\rho}{\hat C}_{mnp}\partial_{\mu}x^{m}\partial_{\nu}x^{n}
\partial_{\rho}x^{p},
\label{S11}
\end{align}
where $x^{m}~(m,n,p=0,1,2,...,10)$ are the 11-dimensional spacetime
coordinates, $\sigma^{\mu}$ are the 3-dimensional world-volume coordinates,
and $g_{mn}$ is 11-dimensional metric.
For a single static flat M2-brane in flat (1+2)-dimensional spacetime
and in an ${\mathbb R}^{8}/{\mathbb Z}_{k}$ orbifold for the transverse space,
we locally have $g_{mn}=\eta_{mn}$. Choosing a static gauge
$\sigma^{\mu}=x^{\mu}$
and regarding the 8 transverse coordinates as the 8 transverse scalar fields
$x^{I}=\lambda X^{I}$, the 3-form field coupling in \eqref{S11}
can be rewritten as
\begin{align}
\frac{\mu_{2}}{3!}\int d^{3}x\epsilon^{\mu\nu\rho}
\left({\hat C}_{\mu\nu\rho}+3\lambda {\hat C}_{\mu\nu I}\partial_{\rho}X^{I}
+3\lambda^{2} {\hat C}_{\mu IJ}\partial_{\nu}X^{I}\partial_{\rho}X^{J}
+\lambda^{3}{\hat C}_{IJK}\partial_{\mu}X^{I}\partial_{\nu}X^{J}
\partial_{\rho}X^{k}
\right).
\label{S11C}
\end{align}
If we compare the obtained action \eqref{S11C} with the action
\eqref{act3} and \eqref{act6} for a single M2-brane,
they look different at first glance. However, we will show that one exactly
coincides with the other.
We can also show that \eqref{S11C} is invariant under
the Abelian gauge transformation
of 3-form field as $C_{(3)}\to C_{(3)}+d\Lambda_{(2)}$.

First of all, if we restrict ourselves to the U(1)$\times$U(1) case of the actions
\eqref{act3} and \eqref{act6}, $\{{\rm Tr}\}$ becomes
trivial due to the Abelian nature.
Noticing that the complex scalar fields commute with each other,
all the terms involving the 6-form field are vanishing.
Therefore the total action with U(1)$\times$U(1) symmetry is obtained only
from the action \eqref{act3}.
Secondly, we note that the action \eqref{act3} involves
U(1)$\times$U(1) gauge fields, while the action \eqref{S11C} does not.
Dynamics of the gauge fields is governed by the Chern-Simons action in
the ABJM theory,
\begin{align}
S_{{\rm CS}}=\int d^{3}x\frac{k}{4\pi}\epsilon^{\mu\nu\rho}
(A_{\mu}\partial_{\nu}A_{\rho}-{\hat A}_{\mu}\partial_{\nu}{\hat A}_{\rho}),
\label{CSt}
\end{align}
and the matter fields couple to the gauge fields only through the covariant
derivative \eqref{cvd}.
If we introduce $A_{\mu}^\pm$ as
\begin{align}
A_{\mu}^\pm=\frac12(A_{\mu}\pm \hat A_{\mu}),
\label{apm}
\end{align}
then the Chern-Simons term \eqref{CSt} is rewritten as
\begin{align}
S_{{\rm CS}}=\int d^{3}x\frac{k}{2\pi}\epsilon^{\mu\nu\rho}A^{+}_{\mu}
F^{-}_{\nu\rho},
\label{CS2}
\end{align}
where $F^{-}_{\mu\nu}=\partial_{\mu}A^{-}_{\nu}
-\partial_{\nu}A^{-}_{\mu}$.
The covariant derivative contains only $A^{-}_{\mu}$,
\begin{align}
D_{\mu}Y^{A}=\partial_{\mu}Y^{A}+2iA^{-}_{\mu}Y^{A}.
\label{Amc}
\end{align}
The U(1) gauge field $A^{+}_{\mu}$ is an auxiliary field which does not couple
to any matter fields and its equation of motion is
$F^{-}_{\mu\nu}=0$.  This makes the other U(1) gauge field $A^{-}_{\mu}$
a pure gauge degree. If we are interested in the physics blind to the
orbifold structure, every covariant derivative can be replaced by an ordinary
partial derivative through a gauge transformation and consequently both
U(1) gauge fields completely decouple from the action \eqref{act3}.

Finally, we see that in \eqref{act3} the transverse scalar fields
$Y^{A}$ are complex, while in \eqref{S11C} the transverse scalar
fields $X^I$ are real. Similarly, the 3-form fields $C_{(3)}$
are complex, while ${\hat C}_{(3)}$ are real.
Therefore, in order to complete the matching between the actions
\eqref{act3} and \eqref{S11C} we relate the complex and real scalar fields
as in \eqref{YA}, which leads to the following identifications among complex and
real 3-form fields:
\begin{align}
C_{\mu\nu\rho}&={\hat C}_{\mu\nu\rho},
\qquad
C_{\mu\nu A}=\frac{1}{2}\big(\hat{C}_{\mu\nu A}-i\hat{C}_{\mu\nu A+4}\big),
\nonumber\\
C_{\mu AB}&= \frac{1}{4}\big(\hat{C}_{\mu AB}-i\hat{C}_{\mu A+4 B}
-i\hat{C}_{\mu A B+4}-\hat{C}_{\mu A+4 B+4}\big),
\nonumber\\
C_{\mu A\bar B}&= \frac{1}{4}\big(\hat{C}_{\mu AB}-i\hat{C}_{\mu A+4 B}
+i\hat{C}_{\mu A B+4}+\hat{C}_{\mu A+4 B+4}\big),
\nonumber\\
C_{ABC}&=
\frac{1}{8}\big({\hat C}_{ABC}-i{\hat C}_{A+4 BC}-i{\hat C}_{A B+4 C}
-i{\hat C}_{AB C+4}
\nonumber \\
&-{\hat C}_{A B+4 C+4}-{\hat C}_{A+4 B C+4}
-{\hat C}_{A+4 B+4 C}+i{\hat C}_{A+4 B+4 C+4}\big),
\nonumber\\
C_{AB\bar C}&=
\frac{1}{8}\big({\hat C}_{ABC}-i{\hat C}_{A+4 BC}-i{\hat C}_{A B+4 C}
+i{\hat C}_{AB C+4}
\nonumber\\
& +{\hat C}_{A B+4 C+4}+{\hat C}_{A+4 B C+4}
-{\hat C}_{A+4 B+4 C}-i{\hat C}_{A+4 B+4 C+4}\big),
\end{align}
and their complex conjugates. Through these identifications, the
action in \eqref{act3} exactly coincides with \eqref{S11C}.
This simple case supports the validity of the proposed actions
in \eqref{act3} and \eqref{act6}, however we need further check for
the general U($N$)$\times$U($N$) case in the subsequent sections.

Note that our result contains that of \cite{Sasaki:2009ij} as a
particular case, where only $C_{\mu A\bar B}$ and its complex conjugate
are taken into account. Since those specific components do not carry
orbifold charge (the charge with respect to $A_{\mu}^{-}$),  they
could be regarded as functions on the orbifold. In fact $C_{(3)}$ is
a differential form, and then each component of it should be
regarded as a section on the orbifold rather than a function. This
allows us  to take into account all components of $C_{(3)}$ even
after orbifolding.

\section{Reduction from M-theory to IIA String Theory}\label{sec4}

In section \ref{sec2} we constructed the WZ-type action, \eqref{act3} and
\eqref{act6}, describing the couplings between the world-volume fields
on multiple M2-branes and the bulk form fields.
Our proposal for these couplings is guided by gauge invariance and
analogy with the coupling of D-branes to the R-R form fields in type
II string theories~\cite{Li:1995pq,Douglas:1995bn,Myers:1999ps}.
However, for the effective theory of multiple M2-branes there
is no known guiding principle on the general form of coupling between the
world-volume fields and the bulk form fields.
On the other hand, in type II
string theories, the coupling between the R-R
form fields and the world-volume fields of D-branes is restricted by consistency
with T-duality.
Therefore,  in order to test the consistency of our proposal, we will
compactify our WZ-type action in \eqref{act3} and \eqref{act6} and then we will compare
 with the corresponding WZ-type action in type IIA string theory.
We first study the simplest case of ABJM theory with U(1)$\times$U(1) gauge
symmetry, and then generalize to the case of U($N$)$\times$U($N$) gauge symmetry.

\subsection{U($1$)$\times$U($1$) gauge symmetry}
\label{ss2-1}

In type IIA string theory the bosonic part of the effective action for
a single D$p$-brane
is\footnote{We put tildes for the fields and the parameters in string theory.}
\begin{align}
S_{10}=S_{{\rm DBI}}+S_{\tilde C}.
\end{align}
$S_{{\rm DBI}}$ is the Abelian DBI action and $S_{\tilde C}$
is the Abelian WZ-type coupling, which is given by~\cite{Myers:1999ps}
\begin{align}
S_{{\tilde C}}=\mu_p\int_{p+1}\left(P\left[\sum {\tilde C}_{(n)}
e^{\tilde{B}}\right]e^{{\tilde\lambda} {\tilde F}}\right),
\label{MCSA1}
\end{align}
where $\mu_p$ is the R-R charge of the D$p$-brane, ${\tilde\lambda}=2\pi
l_{\rm s}^2$ is the string scale, $\tilde{B}$ is the NS-NS 2-form
field, ${\tilde F}=d{\tilde A}$ is field strength of the U(1) gauge
field ${\tilde A}$ of the D-brane, and P[...] is the pull-back
on to the D-brane world-volume. For later convenience we expand
this action explicitly for $p=2$ and ${\tilde C}_{(3)}$, omitting
couplings to other fields,
\begin{align}\label{C3-10}
S_{{\tilde C}}=\mu_2\int \frac{1}{3!} d^3x\epsilon^{\mu\nu\rho}~
\Big(  &{\tilde C}_{\mu\nu\rho} +3\tilde \lambda {\tilde C}_{\mu\nu
i}{\tilde D}_\rho {\tilde X}^i +3\tilde \lambda^2 {\tilde C}_{\mu
ij}{\tilde D}_\nu {\tilde X}^i{\tilde D}_\rho {\tilde X}^j
\nonumber\\
& +\tilde \lambda^3 {\tilde C}_{ijk}{\tilde D}_\mu {\tilde X}^i
{\tilde D}_\nu {\tilde X}^j {\tilde D}_\rho {\tilde X}^k+...\Big),
\end{align}
 where ${\tilde X}^i\,\, (i=1,2,...,7)$ are seven transverse
scalar fields with ${\tilde D}_\mu {\tilde X}^i = \partial_\mu
{\tilde X}^i$ .

We note that in the U($1$)$\times$U($1$) case the scalar fields
$Y^{A}$ commute with each other.
As a result the three commutator $\beta^{AB}_{~C}$ is vanishing and hence
the couplings to the 6-form field are zero.
In addition, the 3-form field should be antisymmetric in all its
global indices.
To compactify one direction transverse to the M2-brane, we apply the
MP Higgsing procedure~\cite{Mukhi:2008ux} to \eqref{act3}.
To that end we turn on a vacuum expectation value (vev) for
the complex scalar $Y^4$ as
\begin{align}
 Y^4=\frac v{2}+ \tilde X^4+i\tilde X^8,
\label{4cp}
\end{align}
while the remaining 3 complex scalars have vanishing vev,
\begin{align}
Y^a=\tilde X^a+i\tilde X^{a+4}, \qquad (a=1,2,3).
\end{align}
Here the 8 real scalar fields, $\tilde X^I=X^I$ $(I=1,...,8)$, stand
for 8 transverse directions. The nonvanishing vev breaks the
U(1)$\times$U(1) gauge symmetry to the diagonal U(1).
Then we take the double scaling limit of
infinite vev $v$ and the Chern-Simons level $k$, keeping $v/k$ finite.
In this limit the gauge field $A^-_\mu$ becomes nondynamical while
$A^+_\mu$ will be a dynamical gauge field.
Using \eqref{apm},  the covariant derivatives of the transverse scalar
fields \eqref{cvd} can be written to the leading order in $v$ as
\begin{align}\label{covab}
D_\mu Y^4=\partial_\mu \tilde X^4+ivA_\mu^-, \qquad
D_\mu Y^a=\partial_\mu \tilde X^a+i\partial_\mu \tilde X^{a+4},
\end{align}
where we have chosen unitary gauge for the field $A^{-}_{\mu}$,
$A_\mu^- -\frac 1v\partial_\mu \tilde X^8 \to A_\mu^-$.

Next we need to relate the 10-dimensional real 3-form fields with
the 11-dimensional complex 3-form fields.
For the first term in \eqref{act3} this relation is obvious and is given by
\begin{align}
\tilde C_{\mu\nu\rho}=C_{\mu\nu\rho}+C^\dagger_{\mu\nu\rho}.
\end{align}
To find the relations for the remaining terms we plug \eqref{covab}
into the action \eqref{act3}.  Then from the second term in
\eqref{act3} we have
\begin{align}
C_{\mu\nu A}D_\rho Y^A+{\rm c.c.} &=(C_{\mu\nu 4}+C^\dagger_{\mu\nu
4})\partial_\rho \tilde X^4+iv(C_{\mu\nu
4}-C^\dagger_{\mu\nu 4})A_\rho^-\nonumber\\
&~~~+(C_{\mu\nu a}+C^\dagger_{\mu\nu a})\partial_\rho \tilde X^a+i(C_{\mu\nu a}
 -C^\dagger_{\mu\nu a})\partial_\rho \tilde X^{a+4}
\nonumber\\
 &=\tilde C_{\mu\nu i}\partial_\rho\tilde
 X^i+v\tilde B_{\mu\nu}A_\rho^-, ~~~~
\end{align}
where we have made the following identification of the R-R 3-form
field and the NS-NS 2-form field in 10-dimensions,
\begin{align}\label{tcb1}
&\tilde C_{\mu\nu a}=(C_{\mu\nu a}+C^\dagger_{\mu\nu a}),
\quad\tilde C_{\mu\nu 4}=(C_{\mu\nu 4}+C^\dagger_{\mu\nu 4}),
\nonumber \\
&\tilde C_{\mu\nu a+4}=i(C_{\mu\nu a}-C^\dagger_{\mu\nu a}),\quad
 \tilde B_{\mu\nu}= i(C_{\mu\nu 4}-C_{\mu\nu\bar 4}).
\end{align}
Similarly, from the third and fourth terms of the action
\eqref{act3} we get
\begin{align}
\lefteqn{ \epsilon^{\mu\nu\rho}\Big(C_{\mu AB}D_\nu Y^A D_\rho
Y^B+C_{\mu A\bar B}D_\nu Y^AD_\rho Y^\dagger_B+{\rm c.c.}\Big) }
\nonumber\\
=&~~\epsilon^{\mu\nu\rho}\Big((-C_{\mu 4\bar a}-C^\dagger_{\mu 4\bar
a}+2C_{\mu a4}+2C^\dagger_{\mu a4}+C_{\mu a\bar4}+C^\dagger_{\mu
a\bar4})\partial_\nu\tilde X^a\partial_\rho\tilde X^4
\nonumber\\
&\hspace{8mm}+i(-C_{\mu 4\bar a}+C^\dagger_{\mu 4\bar a}+2C_{\mu
a4}-2C^\dagger_{\mu a4}+C_{\mu a\bar4}-C^\dagger_{\mu
a\bar4})\partial_\nu\tilde X^4\partial_\rho\tilde X^{a+4}
\nonumber\\
&\hspace{8mm}+iv(C_{\mu 4\bar a}-C^\dagger_{\mu 4\bar a}-2C_{\mu
a4}+2C^\dagger_{\mu a4}+C_{\mu a\bar4}-C^\dagger_{\mu
a\bar4})A_\nu^-\partial_\rho\tilde X^a
\nonumber\\
&\hspace{8mm}+v(C_{\mu 4\bar a}+C^\dagger_{\mu 4\bar a}+2C_{\mu
a4}+2C^\dagger_{\mu a4}-C_{\mu a\bar4}-C^\dagger_{\mu
a\bar4})A_\nu^-\partial_\rho\tilde X^{a+4}
%\nonumber\\
%&\hspace{8mm}
\nonumber\\
&\hspace{8mm}+2iv(C_{\mu 4\bar4}-C^\dagger_{\mu
4\bar4})A_\nu^-\partial_\rho\tilde X^4+(C_{\mu ab}+C^\dagger_{\mu a b}+C_{\mu a\bar
b}+C^\dagger_{\mu a\bar b})\partial_\nu\tilde X^a\partial_\rho\tilde
X^b \nonumber\\
&\hspace{8mm}+i(2C_{\mu ab}-2C^\dagger_{\mu ab}-C_{\mu a \bar
b}+C^\dagger_{\mu a \bar b}-C_{\mu b\bar a}+C^\dagger_{\mu b\bar
a})\partial_\nu\tilde X^a\partial_\rho\tilde X^{b+4}
\nonumber\\
&\hspace{8mm}+(-C_{\mu ab}-C^\dagger_{\mu a b}+C_{\mu a\bar
b}+C^\dagger_{\mu a\bar b})\partial_\nu\tilde
X^{a+4}\partial_\rho\tilde X^{b+4}\Big)\nonumber\\
=&~~\epsilon^{\mu\nu\rho}\Big(\tilde C_{\mu ij}\partial_\nu \tilde
X^i\partial_\rho\tilde X^j+v\tilde B_{\mu i}A_\nu^-\partial_\rho
\tilde X^i\Big). \label{tcb2}
\end{align}
The last step means we have made the identification of
$\tilde C_{\mu ij}$ and $\tilde B_{\mu i}$ as
\begin{align}
&\tilde B_{\mu a}=i(C_{\mu 4\bar a}-C^\dagger_{\mu 4\bar a}-2C_{\mu
a4}+2C^\dagger_{\mu a4}+C_{\mu a\bar4}-C^\dagger_{\mu a\bar4}),
\nonumber\\
&\tilde B_{\mu a+4}=C_{\mu 4\bar a}+C^\dagger_{\mu 4\bar a}+2C_{\mu
a4}+2C^\dagger_{\mu a4}-C_{\mu a\bar4}-C^\dagger_{\mu a\bar4},\nonumber\\
&\tilde B_{\mu 4}=2i(C_{\mu 4\bar4}-C^\dagger_{\mu 4\bar4}),\quad
\tilde C_{\mu ab}=C_{\mu ab}+C^\dagger_{\mu a b}+C_{\mu a\bar
b}+C^\dagger_{\mu a\bar b},\nonumber\\
&\tilde C_{\mu a4}=\frac12(-C_{\mu 4\bar a}-C^\dagger_{\mu 4\bar
a}+2C_{\mu a4}+2C^\dagger_{\mu a4}+C_{\mu a\bar4}+C^\dagger_{\mu
a\bar4}),
\nonumber\\
&\tilde C_{\mu 4a+4}=\frac i2(-C_{\mu 4\bar a}+C^\dagger_{\mu 4\bar
a}+2C_{\mu a4}-2C^\dagger_{\mu a4}+C_{\mu a\bar4}-C^\dagger_{\mu
a\bar4}),
\nonumber\\
&\tilde C_{\mu ab+4}=\frac i2(2C_{\mu ab}-2C^\dagger_{\mu ab}-C_{\mu
a \bar b}+C^\dagger_{\mu a \bar b}-C_{\mu b\bar a}+C^\dagger_{\mu
b\bar a}),
\nonumber\\
&\tilde C_{\mu a+4b+4}=-C_{\mu ab}-C^\dagger_{\mu a b}+C_{\mu a\bar
b}+C^\dagger_{\mu a\bar b}.
\end{align}

For the last two terms  of the action \eqref{act3}, we have
\begin{align}
\lefteqn{ \epsilon^{\mu\nu\rho}\Big(C_{ABC}D_\mu Y^AD_\nu Y^BD_\rho
Y^C +C_{AB\bar C}D_\mu Y^AD_\nu Y^BD_\rho Y^\dagger_C \Big)+{\rm
c.c.}}\nonumber\\
 =&\epsilon^{\mu\nu\rho}\Big(\tilde
C_{ijk}\partial_\mu\tilde X^{i}
\partial_\nu\tilde X^{j}\partial_\rho\tilde
X^{k}+v\tilde B_{ij}A_\mu^-\partial_\nu\tilde
X^{i}\partial_\rho\tilde X^{j}\Big). \label{tcb6}
\end{align}
where as usual we made the following identification of the 10-dimensional form fields
\begin{align}
&\tilde B_{a4}=-2i(C_{a4\bar 4}-C^\dagger_{a4\bar 4}),~~~~\tilde
B_{4a+4}=-2(C_{a4\bar 4}+C^\dagger_{a4\bar 4}),
\nonumber\\
&\tilde B_{ab}=i(3C_{ab4}-3C^\dagger_{ab4}+2C_{a\bar
b4}-2C^\dagger_{a\bar b4}-C_{ab\bar4} +C^\dagger_{ab\bar 4}),
\nonumber\\
&\tilde B_{ab+4}=-3C_{4ab}-3C^\dagger_{ab4} +C_{ab\bar
4}+C^\dagger_{ab\bar 4}+C_{a\bar b 4}+C^\dagger_{a\bar b 4}+C_{b\bar
a4} +C^\dagger_{b\bar a 4},
\nonumber\\
&\tilde B_{a+4b+4}=i(-3C_{ab4}+3C^\dagger_{ab4}+2C_{a\bar
b4}-2C^\dagger_{a\bar b4}+C_{ab\bar4} -C^\dagger_{ab\bar 4}),
\nonumber\\
&\tilde C_{ab4}=\frac13(3C_{ab4}+3C^\dagger_{ab4}+2C_{a\bar
b4}+2C^\dagger_{a\bar b4}+C_{ab\bar4} +C^\dagger_{ab\bar 4}),
\nonumber\\
&\tilde C_{a4b+4}=\frac i3(-3C_{4ab}+3C^\dagger_{ab4} -C_{ab\bar
4}+C^\dagger_{ab\bar 4}+C_{a\bar b 4}-C^\dagger_{a\bar b 4}+C_{b\bar
a4} -C^\dagger_{b\bar a 4}),
\nonumber\\
&\tilde C_{4a+4b+4}=\frac13(-3C_{ab4}-3C^\dagger_{ab4}+2C_{a\bar
b4}+2C^\dagger_{a\bar b4}-C_{ab\bar4} -C^\dagger_{ab\bar 4}),
\nonumber\\
&\tilde C_{abc}=(C_{abc}+C^\dagger_{abc}+C_{ab\bar c}
+C^\dagger_{ab\bar c}),\nonumber\\
&\tilde C_{abc+4}=\frac i3(3C_{abc} -3C^\dagger_{abc}-C_{ab\bar c}
+C^\dagger_{ab\bar c}-C_{ac\bar b} +C^\dagger_{ac\bar b}-C_{cb\bar
a} +C^\dagger_{cb\bar a}),
\nonumber\\
&\tilde C_{ab+4c+4}=\frac 13(-3C_{abc} -3C^\dagger_{abc}+2C_{ab\bar
c} +2C^\dagger_{ab\bar c}+C_{cb\bar a} +C^\dagger_{cb\bar a}), \nonumber\\
&\tilde C_{a+4b+4c+4}=i(-C_{abc} +C^\dagger_{abc}+C_{ab\bar
c}-C^\dagger_{ab\bar c}).
\end{align}

Collecting all the terms, we obtain
\begin{align}\label{acu1-10}
S_{\tilde C}=\mu_2\int d^3x\, \frac{1}{3!}
\epsilon^{\mu\nu\rho}&\Big[\tilde C_{\mu\nu\rho}+3\lambda(\tilde
C_{\mu\nu i}\partial_\rho\tilde X^i+v\tilde
B_{\mu\nu}A_\rho^-)+3\lambda^2(\tilde C_{\mu ij}\partial_\nu \tilde
X^i\partial_\rho\tilde X^j+v\tilde B_{\mu i}A_\nu^-\partial_\rho
\tilde X^i)\nonumber\\
&~~+\lambda^3(\tilde C_{ijk}\partial_\mu\tilde X^{i}
\partial_\nu\tilde X^{j}\partial_\rho\tilde
X^{k}+v\tilde B_{ij}A_\mu^-\partial_\nu\tilde
X^{i}\partial_\rho\tilde X^{j})\Big].
\end{align}
Next we integrate out the auxiliary field $A^-_{\mu}$.
In order to do so, we should take into account the bosonic part of the original
U(1)$\times$U(1) ABJM action. After the MP compactification procedure,
the leading order in $v$ of the ABJM action takes the following form,
\begin{align}
\int d^3x \, \Big(-\tilde D_\mu \tilde X^i \tilde
D^\mu \tilde X^i-v^2A_\mu^-A^{-\mu}+\frac k{2\pi}\
\epsilon^{\mu\nu\rho}A^-_\mu\tilde F_{\nu\rho}\Big)+{\cal
O}\big(\frac 1v\big),
\label{CS+X}
\end{align}
where ${\tilde F}_{\mu\nu}=\partial_\mu A^+_\nu-\partial_\nu A^+_\mu.$
The equation of motion for the auxiliary field  $A^-_{\mu}$ can be
obtained from the variation of \eqref{acu1-10} and \eqref{CS+X}.
Solving this equation of motion to the leading order in $v$ expresses
the auxiliary fields in terms of dynamical fields as
\begin{align}
A^{-}_\mu =\frac k{4\pi v^2}\epsilon_\mu\!\!~^{\nu\rho}\Big( {\tilde
F}_{\nu\rho}+{\mu_2 v\lambda}\frac{2\pi}k P[\tilde
B_{\nu\rho}]\Big)=\frac{1}{2g_{{\rm YM}} v}\,\epsilon_\mu^{~\nu\rho}
\Big({ \tilde F}_{\nu\rho}+\frac{1}{\tilde\lambda} P[\tilde
B_{\nu\rho}]\Big). \label{Amu-}
\end{align}
where
\begin{align}\label{PBmunu}
&P[\tilde B_{\mu\nu}]=\frac12\Big(\tilde B_{\mu\nu}+\lambda \tilde
B_{\mu i} \partial_\nu \tilde X^i+\frac{\lambda^2}3\tilde B_{ij}
\partial_\nu \tilde X^i\partial_\rho \tilde X^j\Big),
\nonumber \\
&g_{{\rm YM}}=\frac {2\pi v}k,\qquad g_{{\rm
s}}=g^2_{{\rm YM}} l_{{\rm s}}, \qquad \tilde\lambda=2\pi l_{{\rm s}}^2
\end{align}
with $\mu_{2}\lambda=1/{2\pi l_{\rm P}^{3/2}}$
and $l_{\rm P}=g_{{\rm s}}^{1/3}l_{{\rm s}}$.
For dimensional reason we also rescale the scalar fields as
$\tilde X^i\to \frac {\tilde X^i}{g_{\rm YM}}$. Substituting  $A_\mu^-$ into
the actions (\ref{acu1-10}) and (\ref{CS+X}), and rearranging the terms, we obtain
\begin{align}
S_{10}=\int d^3x\Bigg\{\, \frac{\mu_2}{3!}
\epsilon^{\mu\nu\rho}&\Big[\tilde C_{\mu\nu\rho}+3\tilde
\lambda\tilde C_{\mu\nu i}\partial_\rho\tilde X^i+3\tilde
\lambda^2\tilde C_{\mu ij}\partial_\nu \tilde X^i\partial_\rho\tilde
X^j+\tilde \lambda^3\tilde C_{ijk}\partial_\mu\tilde X^{i}
\partial_\nu\tilde X^{j}\partial_\rho\tilde
X^{k}\Big]\nonumber\\
&+\frac1{g^2_{{\rm YM}}}\Big[-\tilde D_\mu \tilde X^i \tilde D^\mu
\tilde X^i-\frac12\Big({ \tilde F}_{\mu\nu}+\frac{1}{\tilde\lambda}
P[\tilde B_{\mu\nu}]\Big)^2\Big]\Bigg\}.
\label{acu1-10-2}
\end{align}
As anticipated, the reduction to 10-dimensions results in the linearized
DBI action and the coupling of the R-R 3-form field to D2-brane
\eqref{C3-10}.

\subsection{U($N$)$\times$U($N$) gauge symmetry}\label{ss2-2}

In the pervious section we have verified that, when compactified to
10-dimensions, the {U(1)$\times$U(1) ABJM theory coupled to the
11-dimensional 3-form field gives rise to the theory of a single
D2-brane coupled to the 10-dimensional R-R 3-form field and the
NS-NS 2-form field. In this section we will extend this procedure to
the case of non-Abelian gauge symmetry. In particular we will
consider the U($N$)$\times$U($N$) ABJM theory coupled to the 11-dimensional
3-form and 6-form fields.
In string theory,
the non-Abelian extension of \eqref{MCSA1} is given in \cite{Myers:1999ps} as
\begin{align}
S_{{\tilde C}}=\mu_p\int_{p+1}\mathrm{Tr}\left(P\left[
e^{i\tilde{\lambda}\mathbf{i}_{\tilde{X}}^{2}}\sum {\tilde C}_{(n)}
e^{\tilde{B}}\right]e^{{\tilde\lambda} {\tilde F}}\right),
\label{MCSA2}
\end{align}
where $\mathbf{i}_{\tilde{X}}$ denotes the contraction (interior product)
with $\tilde{X}^{i}$.
Note that since $\tilde{X}^{i}$ is now the $N\times N$ matrix,
$\mathbf{i}_{\tilde{X}}^{2}$ is nonzero
and given in terms of the commutator of $\tilde{X}^{i}$.

For clarity of presentation we treat the terms involving the 3-
and 6-form fields separately.
The reduction to 10-dimension is achieved by breaking the
U($N$)$\times$U($N$) gauge symmetry down to U($N$), and
the scalar fields are in the adjoint representation
of the unbroken U($N$). Therefore, the transverse scalars $X^I$
introduced in \eqref{YA} can be split into its trace and traceless part as
\begin{align}
&X^I={\check X}^I+i{\hat X}^{I}={\check X}_0^IT^0+i{\hat
X}_\alpha^{I}T^\alpha ,
\end{align}
where $T^0$ and $T^\alpha ~~ (\alpha=1,..., N^2-1$) are the generators of
the unbroken U($N$).
Then the covariant derivative \eqref{cvd} becomes
\begin{align}
D_\mu Y^A=\tilde D_\mu X^A+i\tilde D_\mu
X^{A+4}+i\{A_\mu^-,X^A+iX^{A+4}\},
\label{cv2}
\end{align}
where we used non-Abelian version of \eqref{apm} and $\tilde
D_\mu X=\partial_\mu X+i[A_\mu^+,X]$.

Now let us turn on vev for the trace part of $Y^4$ as
\begin{align}
Y^4=\frac v2 T^0 +X^4+iX^{8},
\end{align}
and introduce 7 Hermitian scalars fields in the adjoint representation
of the U($N$):
\begin{align}
\tilde X^4={\check X}^4-{\hat X}^{8},\quad
 \tilde X^{a}={\check
X}^{a}-{\hat X}^{a+4},\quad
 \tilde X^{a+4}={\check
X}^{a+4}+{\hat X}^{a},\quad (a=1,2,3).
\end{align}
In the double scaling limit of infinite vev $v$ and the Chern-Simons level $k$,
the covariant derivative \eqref{cv2} to the leading order in $v$ becomes
\begin{align}
D_\mu Y^4&=\tilde D_\mu ({\check X}^4-{\hat X}^{8})+iv[A_\mu^-+\frac
1v(\tilde D_\mu ({\check X}^8+{\hat X}^{4}))] =\tilde D_\mu \tilde
X^4+ivA_\mu^- ,
\nonumber\\
D_\mu Y^a&=\tilde D_\mu [({\check X}^a+i{\hat X}^{a})+i({\check
X}^{a+4}+i{\hat X}^{a+4})] =\tilde D_\mu \tilde X^{a}+i\tilde
D_\mu\tilde X^{a+4}, \label{covder}\end{align}
 where we have made
a gauge choice where  $A_\mu^-\to A_\mu^--\frac 1v \tilde D_\mu ({\check
X}^8+{\hat X}^{4})$.
In the same limit, the three commutator terms $\beta^{AB}_{~C}$ are reduced to
\begin{align}\label{redbeta}
\beta^{a4}_{~4} &= \frac{v}{2}\big([\tilde X^a,\,\tilde X^4]
+ i [\tilde X^{a+4},\, \tilde X^4]\big),
\nonumber \\
\beta^{ab}_{~4} &= \frac{v}{4}\big([\tilde X^a,\, \tilde X^b]
+ i[\tilde X^a,\, \tilde X^{b+4}]+ i[\tilde X^{a+4},\,\tilde X^b]
- [\tilde X^{a+4},\, \tilde X^{b+4}]\big),
\nonumber \\
\beta^{a4}_{~b} &= \frac{v}{4}\big([\tilde X^a,\, \tilde X^b]
- i[\tilde X^a,\, \tilde X^{b+4}]+ i[\tilde X^{a+4},\,\tilde X^b]
+ [\tilde X^{a+4},\, \tilde X^{b+4}]\big).
\end{align}

\subsubsection{3-form fields}

As has been done in the previous subsection, we calculate each term
in the WZ-type action \eqref{act3} in the compactification limit,
and identify it with the corresponding WZ-type coupling in type IIA
string theory. The first term in \eqref{act3} can simply be identified
with the corresponding 3-form field in type IIA string theory as
\begin{align}\label{C3-1}
\tilde C_{\mu\nu\rho}=C_{\mu\nu\rho}+C^{\dagger}_{\mu\nu\rho}.
\end{align}
Using \eqref{covder}, from the second term in \eqref{act3} we obtain
\begin{align}\label{C2-4}
C_{\mu\nu A} D_\rho Y^A +{\rm c.c.}=\tilde C_{\mu\nu i}D_\rho
\tilde X^{i}+v\tilde B_{\mu\nu}A_\rho^-,
\end{align}
where we have identified the 3-form fields in M-theory with the R-R
and NS-NS form fields in IIA string theory as
\begin{align}
\tilde C_{\mu\nu {a+4}}=&~i(C_{\mu\nu a}-C^\dagger_{\mu\nu a}),
\qquad \tilde C_{\mu\nu a}=C_{\mu\nu a}+C^\dagger_{\mu\nu a},
\nonumber\\
\tilde C_{\mu\nu 4}=&~C_{\mu\nu 4}+C^\dagger_{\mu\nu4},\qquad
~~~~\tilde B_{\mu\nu}=i(C_{\mu\nu 4}-C^\dagger_{\mu\nu4}).
\end{align}

For the remaining terms in \eqref{act3} the reduction to 10-dimensions seems
more subtle. As an example let us consider the following
particular term:
\begin{align}
\{{\rm Tr}\}(C_{\mu A\bar B} D_\nu Y^A D_\rho Y^\dagger_B)=(C_{\mu A\bar B})^{\hat
ab}_{a\hat b} (D_\nu Y^A)^{a}_{\hat a} (D_\rho Y^\dagger_B)^{\hat
b}_b.
\end{align}
After the symmetry breaking the complex scalar fields and
their complex conjugates are all in the adjoint representation of
the unbroken U($N$). Therefore, the hatted and unhatted gauge indices
are indistinguishable so that we may  write
\begin{align}\label{C2-1}
\{{\rm Tr}\}(C_{\mu A\bar B} D_\nu Y^A D_\rho Y^\dagger_B)=(C_{\mu A\bar B})_{ab}^{cd}
(D_\nu Y^A)^{a}_{c} (D_\rho Y^\dagger_B)^{ b}_d.
\end{align}
Based on the Taylor expansion in \eqref{B2}, in the compactification limit,
the leading order terms of $(C_{\mu A\bar B})_{ab}^{cd}$ are given by
 \begin{align}
(C_{\mu A\bar B})_{ab}^{cd}=\delta^d_a(C^{(1)}_{\mu A\bar B})^{c}_{b}
+\delta^c_b(C^{(2)}_{\mu A\bar B})^{d}_{a},
\end{align}
where $C^{(1)}$ and $C^{(2)}$ can depend on worldvolume coordinates
as well as transverse scalar
fields.\footnote{See Appendix \ref{appA} for clarifications.} As a
result the WZ-type coupling in (\ref{C2-1}) gives
\begin{align}\label{C2-3}
\{{\rm Tr}\}(C_{\mu A\bar B} D_\nu Y^A D_\rho Y^\dagger_B)
={\rm Tr}(C^{(1)}_{\mu A\bar B}D_\nu Y^AD_\rho Y^\dagger_B)
+{\rm Tr}(C^{(2)}_{\mu A\bar B}D_\rho Y^\dagger_BD_\nu Y^A),
\end{align}
where `Tr' on the right-hand side represents the ordinary trace of
$N\times N$ matrices.
From the WZ-type coupling in \eqref{C2-1}, we get the two types of terms
in \eqref{C2-3} and have introduced two types of form fields,
$C^{(1)}_{\mu A\bar B}$ and $C^{(2)}_{\mu A\bar B}$.
This is true for the other terms
involving both $D_\mu Y^A$ and $D_\mu Y^\dagger_A$. However, for
the terms involving only $D_\mu Y^A$ or only $D_\mu Y^\dagger_A$, we
have only one type of term. For instance
\begin{align}
\epsilon^{\mu\nu\rho}\{{\rm Tr}\}(C_{\mu A B} D_\nu Y^A D_\rho
Y^B)&=\epsilon^{\mu\nu\rho}{\rm Tr}\big(\hat C^{(1)}_{\mu AB}D_\nu Y^AD_\rho
Y^B+\hat C^{(2)}_{\mu AB}D_\rho Y^BD_\nu Y^A\big)
\nonumber\\
&=\epsilon^{\mu\nu\rho}{\rm Tr}(C^{(3)}_{\mu AB}D_\nu Y^AD_\rho Y^B),
\end{align}
where $C^{(3)}_{\mu AB}=\hat C^{(1)}_{\mu AB}-\hat C^{(2)}_{\mu BA}$.

Using this procedure, from the third and
the fourth terms in \eqref{act3} we have
\begin{align}\label{2te}
\epsilon^{\mu\nu\rho}\{{\rm Tr}\}\big(&C_{\mu A\bar B} D_\nu Y^A D_\rho
Y^\dagger_B+C_{\mu AB} D_\nu Y^A D_\rho Y^B\big)+({\rm c.c.})
\\
=\epsilon^{\mu\nu\rho}&{\rm Tr}\big(C^{(1)}_{\mu A\bar B}D_\nu Y^A D_\rho
Y^\dagger_B-C^{(2)}_{\mu A\bar B}D_\nu Y^\dagger_BD_\rho Y^A
+C^{(3)}_{\mu AB} D_\nu Y^A D_\rho Y^B\big)+{\rm (c.c.)}\nonumber\\
=\epsilon^{\mu\nu\rho}&{\rm Tr}\big(C^{(1)}_{\mu 4\bar 4}D_\nu Y^4 D_\rho
Y^\dagger_4- C^{(2)}_{\mu 4\bar 4}D_\nu Y^\dagger_4D_\rho
Y^4+C^{(3)}_{\mu 44}D_\nu Y^4D_\rho Y^4+C^{(1)}_{\mu 4\bar a} D_\nu
Y^4 D_\rho Y^\dagger_a
\nonumber\\
&+C^{(1)}_{\mu a\bar 4} D_\nu Y^a D_\rho Y^\dagger_4-C^{(2)}_{\mu
a\bar 4} D_\nu Y^\dagger_4 D_\rho Y^a-C^{(2)}_{\mu 4\bar a} D_\nu
Y^\dagger_a D_\rho Y^4+C^{(3)}_{\mu 4 a}D_\nu Y^4D_\rho Y^a
\nonumber\\
& +C^{(3)}_{\mu a 4}D_\nu
Y^aD_\rho Y^4+C^{(1)}_{\mu a\bar b}D_\nu Y^a D_\rho
Y^\dagger_b-C^{(2)}_{\mu a\bar b}D_\nu Y^\dagger_b D_\rho Y^a
+C^{(3)}_{\mu ab}D_\nu Y^a D_\rho Y^b\big)+{\rm (c.c.)}.
\nonumber
\end{align}
 Then we use (\ref{covder}) to replace the covariant derivatives in
 (\ref{2te}). Unlike \eqref{C2-4}, this replacement does not exactly
 produce the expected WZ-type coupling in type IIA string theory.
 In general, when we plug (\ref{covder}) into (\ref{act3}), in addition
 to the expected 10-dimensional WZ-type couplings, it generates some unknown
 terms in string theory, involving quadratic or higher order in either
 $A_\mu^-$ or $\tilde D_\mu\tilde X^4$.
 To reproduce all the known correct couplings from (\ref{2te}), we have to impose
 some constraints on the different types of 3-form fields which we have
 introduced.  These constraints are obtained by requiring that the NS-NS
 and R-R form fields obtained from the reduction to 10-dimensions should
  be antisymmetric and the coefficients of the unknown terms involving
  $A^-_{\mu}A^-_{\nu}$ and $\tilde D_{\mu}\tilde X^4\tilde D_{\nu}\tilde X^4$
  should vanish.  In order to get a symmetrized 10-dimensional products
  we also require that the coefficients of $A^-_{\mu}\tilde D_{\nu}\tilde X^i$ and
  $\tilde D_{\nu}\tilde X^iA^-_{\mu}$ are equal. From these conditions we obtain
   the following constraints:
\begin{align}
&C_{\mu A\bar B}^{(1)} - C_{\mu B\bar A}^{(1)\dagger} =
C_{\mu A\bar B}^{(2)} - C_{\mu B\bar A}^{(2)\dagger},
\quad C^{(3)}_{\mu AB} = C^{(3)}_{\mu [AB]},
\qquad (A,B=1,2,3,4),
\end{align}
where the notation $[AB...]$ denotes the antisymmetrization of indices.
With these constraints (\ref{2te}) is reduced to
\begin{align}\label{C3-2}
\epsilon^{\mu\nu\rho}&\{{\rm Tr}\}\big(C_{\mu A\bar B} D_\nu Y^A D_\rho
Y^\dagger_B+C_{\mu AB} D_\nu Y^A D_\rho Y^B\big)+{\rm (c.c.)}
\nonumber\\
&=~~\epsilon^{\mu\nu\rho}{\rm Tr}\big(\tilde C_{\mu ij}
\tilde D_\nu\tilde X^i\tilde D_\rho\tilde X^j
+v\tilde B_{\mu i}\langle\hspace{-0.7mm}\langle
A_\nu^-\tilde D_\rho\tilde X^i \rangle\hspace{-0.7mm}\rangle\big),
\end{align}
where the R-R form fields $\tilde C_{\mu ij}$ and the NS-NS form fields
$\tilde B_{\mu i}$ are identified as
\begin{align}
&\tilde B_{\mu4}=4i\big(C^{(1)}_{\mu4\bar4}-
C^{(1)\dagger}_{\mu 4\bar 4}\big),
\nonumber\\
&\tilde B_{\mu a}=2i\big(C^{(1)}_{\mu 4\bar a}-C^{(1)\dagger}_{\mu 4\bar
a}+ C^{(1)}_{\mu a\bar4}-C^{(1)\dagger}_{\mu a\bar 4}+C^{(3)}_{\mu 4a}
-C^{(3)\dagger}_{\mu 4a}\big),
\nonumber\\
&\tilde B_{\mu a+4}=2\big(C^{(1)}_{\mu 4\bar a}+C^{(1)\dagger}_{\mu
4\bar a}- C^{(1)}_{\mu a\bar4}-C^{(1)\dagger}_{\mu a\bar 4}
-C^{(3)}_{\mu 4a} -C^{(3)\dagger}_{\mu 4a}\big),
\nonumber\\
&\tilde C_{\mu4a}= C^{(1)}_{\mu 4\bar a}
+C^{(1)\dagger}_{\mu 4\bar a}- C^{(1)}_{\mu a\bar4}-
C^{(1)\dagger}_{\mu a\bar 4} +C^{(3)}_{\mu 4a} +C^{(3)\dagger}_{\mu 4a},
\nonumber\\
&\tilde C_{\mu4a+4}=-i\big(C^{(1)}_{\mu 4\bar a}
-C^{(1)\dagger}_{\mu 4\bar a}+ C^{(1)}_{\mu a\bar4}
-C^{(1)\dagger}_{\mu a\bar 4}-C^{(3)}_{\mu 4a} +C^{(3)\dagger}_{\mu 4a}\big),
\nonumber\\
&\tilde C_{\mu ab}=C^{(1)\dagger}_{\mu a\bar b}-C^{(1)}_{\mu b\bar
a}+ C^{(2)}_{\mu a\bar b}-C^{(2)\dagger}_{\mu b\bar a}-C^{(3)\dagger}_{\mu ab}
+C^{(3)}_{\mu ab},
\nonumber\\
&\tilde C_{\mu ab+4}
=i\big(C^{(1)\dagger}_{\mu a\bar b}-C^{(1)}_{\mu b\bar a}
-C^{(2)}_{\mu a\bar b}+C^{(2)\dagger}_{b\bar a}+C^{(3)\dagger}_{\mu ab}
-C^{(3)}_{\mu ba}\big),
\nonumber\\
&\tilde C_{\mu a+4 b+4}=C^{(1)\dagger}_{\mu a\bar b}-C^{(1)}_{\mu
b\bar a}+ C^{(2)}_{\mu a\bar b}-C^{(2)\dagger}_{\mu b\bar a}
+C^{(3)\dagger}_{\mu a b} -C^{(3)}_{\mu ab}.
\end{align}
For simplicity we used a short hand notation
$\langle\hspace{-0.7mm}\langle ...\rangle\hspace{-0.7mm}\rangle$ for
symmetrized product in \eqref{C3-2}. For instance,
\begin{align}
\langle\hspace{-0.7mm}\langle A_\nu^-\tilde
D_\rho\tilde X^i\rangle\hspace{-0.7mm}\rangle
=\frac12\big(A_\nu^-\tilde
D_\rho\tilde X^i+\tilde D_\rho\tilde X^i A_\nu^-\big).
\end{align}

Next we consider the last two terms in (\ref{act3}).
After the breaking of the gauge symmetry these terms can
be written as
\begin{align}\label{C34}
\epsilon^{\mu\nu\rho}&\{{\rm Tr}\}\big(C_{ABC} D_\mu Y^A D_\nu Y^BD_\rho
Y^C+C_{AB\bar C} D_\mu Y^A
D_\nu Y^BD_\rho Y^\dagger_C\big)+{\rm (c.c.)}\nonumber\\
=\epsilon^{\mu\nu\rho}&{\rm Tr}\big(C^{(1)}_{ABC}
 D_\mu Y^A D_\nu Y^BD_\rho
Y^C+C^{(2)}_{AB\bar C} D_\mu Y^A D_\nu Y^BD_\rho
Y^\dagger_C-C^{(3)}_{AC\bar B} D_\mu Y^A D_\nu Y^\dagger_BD_\rho
Y^C\nonumber\\
&+C^{(4)}_{BC\bar A} D_\mu Y^\dagger_A
D_\nu Y^BD_\rho Y^C\big)+{\rm (c.c.)}.\nonumber\\
=\epsilon^{\mu\nu\rho}&{\rm Tr}\big(C^{(1)}_{44a}
 D_\mu Y^4 D_\nu Y^4D_\rho
Y^a+C^{(1)}_{4a4}
 D_\mu Y^4 D_\nu Y^aD_\rho
Y^4+C^{(1)}_{a44}
 D_\mu Y^a D_\nu Y^4D_\rho
Y^4\nonumber\\
&+C^{(2)}_{44\bar a} D_\mu Y^4 D_\nu Y^4D_\rho
Y^\dagger_a+C^{(2)}_{4a\bar 4} D_\mu Y^4 D_\nu Y^aD_\rho
Y^\dagger_4+C^{(2)}_{a4\bar 4} D_\mu Y^a
D_\nu Y^4D_\rho Y^\dagger_4\nonumber\\
&-C^{(3)}_{4a\bar 4} D_\mu Y^4 D_\nu Y^\dagger_4D_\rho
Y^a-C^{(3)}_{44\bar a} D_\mu Y^4 D_\nu Y^\dagger_aD_\rho
Y^4-C^{(3)}_{a4\bar 4} D_\mu Y^a D_\nu Y^\dagger_4D_\rho
Y^4\nonumber\\
&+C^{(4)}_{4a\bar 4} D_\mu Y^\dagger_4 D_\nu Y^4D_\rho
Y^a+C^{(4)}_{a4\bar 4} D_\mu Y^\dagger_4 D_\nu Y^aD_\rho
Y^4+C^{(4)}_{44\bar a} D_\mu Y^\dagger_a
D_\nu Y^4D_\rho Y^4\nonumber\\
&+C^{(1)}_{4ab}
 D_\mu Y^4 D_\nu Y^aD_\rho
Y^b+C^{(1)}_{a4b}
 D_\mu Y^a D_\nu Y^4D_\rho
Y^b+C^{(1)}_{ab4}
 D_\mu Y^a D_\nu Y^bD_\rho
Y^4\nonumber\\
&+C^{(2)}_{4a\bar b} D_\mu Y^4 D_\nu Y^aD_\rho
Y^\dagger_b+C^{(2)}_{a4\bar b} D_\mu Y^a D_\nu Y^4D_\rho
Y^\dagger_b+C^{(2)}_{ab\bar 4} D_\mu Y^a
D_\nu Y^bD_\rho Y^\dagger_4\nonumber\\
&-C^{(3)}_{4b\bar a} D_\mu Y^4 D_\nu Y^\dagger_aD_\rho
Y^b-C^{(3)}_{ab\bar 4} D_\mu Y^a D_\nu Y^\dagger_4D_\rho
Y^b-C^{(3)}_{a4\bar b} D_\mu Y^a D_\nu Y^\dagger_bD_\rho
Y^4\nonumber\\
&+C^{(4)}_{ab\bar 4} D_\mu Y^\dagger_4 D_\nu Y^aD_\rho
Y^b+C^{(4)}_{4b\bar a} D_\mu Y^\dagger_a D_\nu Y^4D_\rho
Y^b+C^{(4)}_{b4\bar a} D_\mu Y^\dagger_a
D_\nu Y^bD_\rho Y^4\nonumber\\
&+C^{(1)}_{abc}
 D_\mu Y^a D_\nu Y^bD_\rho
Y^c+C^{(2)}_{ab\bar c} D_\mu Y^a D_\nu Y^bD_\rho
Y^\dagger_c-C^{(3)}_{ac\bar b} D_\mu Y^a D_\nu Y^\dagger_bD_\rho
Y^c\nonumber\\
&+C^{(4)}_{bc\bar a} D_\mu Y^\dagger_a
D_\nu Y^bD_\rho Y^c\big)+{\rm (c.c.)}.
\end{align}
In the last step we assumed the terms involving $C_{444}$ and $C_{44\bar4}$
are vanishing. This is because, as we see from \eqref{covder}, those terms
give rise to only terms that are quadratic or higher order in $A_{\mu}^-$
or $\tilde D_\mu\tilde X^4$, which are not allowed in string theory.

The next step is  to replace the covariant derivatives
in \eqref{C34} by using (\ref{covder}).
As we did in \eqref{2te}, we require that the R-R and NS-NS
form fields which are obtained from the reduction to 10-dimensions should
be antisymmetric in interchange of indices and the coefficients of the unwanted terms involving
$A^-_{\mu}A^-_{\nu}\tilde D_\rho\tilde X^{a}$ or
$\tilde D_{\mu}\tilde X^4\tilde D_{\nu}\tilde X^4\tilde D_{\rho}\tilde X^a$ vanish.
We also impose more conditions to produce
symmetrized matrix products in 10-dimensions.
For instance we require that the coefficients of
$(A^-_\mu \tilde D_\nu\tilde X^a\tilde D_\rho\tilde X^b)$,
$(\tilde D_\nu\tilde X^aA^-_\mu \tilde D_\rho\tilde X^b)$, and
$(\tilde D_\nu\tilde X^a\tilde D_\rho\tilde X^bA^-_\mu )$ are equal.
The constraints from these requirements are
\begin{align}
C^{(1)}_{ABC} =C^{(1)}_{[ABC]}, \quad C^{(2)}_{AB\bar C}
=C^{(2)}_{[AB]\bar C}=C^{(3)}_{AB\bar C} =C^{(3)}_{[AB]\bar C}
= C^{(4)}_{AB\bar C} =C^{(4)}_{[AB]\bar C}.
\end{align}
With these constraints we get
\begin{align}\label{C3-3}
\epsilon^{\mu\nu\rho}&\{{\rm Tr}\}\big(C_{ABC} D_\mu Y^A D_\nu Y^BD_\rho
Y^C+C_{AB\bar C} D_\mu Y^A D_\nu Y^BD_\rho Y^\dagger_C\big)+{\rm (c.c.)}
\nonumber\\
=&~\epsilon^{\mu\nu\rho}{\rm Tr}\big(v\tilde
B_{ij}\langle\hspace{-0.7mm}\langle A^-_\mu
\tilde D_\nu \tilde X^i\tilde D_\rho \tilde X^j\rangle\hspace{-0.7mm}\rangle+\tilde
C_{ijk}\tilde D_\mu \tilde X^i\tilde D_\nu \tilde X^j
 \tilde D_\rho \tilde X^k\big),
\end{align}
where
\begin{align}
&\tilde B_{a4}=-3i\big(C^{(4)}_{a4\bar 4}-C^{(4)\dagger}_{a4\bar 4}\big), \quad
\tilde B_{4a+4}= -3\big(C^{(4)}_{a4\bar 4}+C^{(4)\dagger}_{a4\bar 4}\big),
\nonumber\\
&\tilde B_{ab} = 3i \big(C^{(1)}_{ab4}-C^{(1)\dagger}_{ab4}
-C^{(4)}_{ab\bar 4}+ C^{(4)\dagger}_{ab\bar 4}
+C^{(4)}_{b4\bar a}- C^{(4)\dagger}_{ b4\bar a}
-C^{(4)}_{a4\bar b} +C^{(4)\dagger}_{a4\bar b}\big),
\nonumber\\
&\tilde B_{ab+4} = -3 \big(
C^{(1)}_{ab4}+C^{(1)\dagger}_{ab4}
-C^{(4)}_{ab\bar 4}- C^{(4)\dagger}_{ab\bar 4}
+C^{(4)}_{b4\bar a}+ C^{(4)\dagger}_{ b4\bar a}
+C^{(4)}_{a4\bar b} +C^{(4)\dagger}_{a4\bar b}\big),
\nonumber \\
&\tilde B_{a+4b+4} = -3i \big(C^{(1)}_{ab4}-C^{(1)\dagger}_{ab4}
-C^{(4)}_{ab\bar 4}+ C^{(4)\dagger}_{ab\bar 4}
-C^{(4)}_{b4\bar a}+ C^{(4)\dagger}_{ b4\bar a}
+C^{(4)}_{a4\bar b} -C^{(4)\dagger}_{a4\bar b}\big),
\nonumber \\
&\tilde C_{ab4} = 3 \big(
C^{(1)}_{ab4}+C^{(1)\dagger}_{ab4}
+C^{(4)}_{ab\bar 4}+ C^{(4)\dagger}_{ab\bar 4}
+C^{(4)}_{b4\bar a}+ C^{(4)\dagger}_{ b4\bar a}
-C^{(4)}_{a4\bar b} -C^{(4)\dagger}_{a4\bar b}\big),
\nonumber \\
&\tilde C_{a4b+4} = -3 \big(
C^{(1)}_{ab4}-C^{(1)\dagger}_{ab4}
+C^{(4)}_{ab\bar 4}- C^{(4)\dagger}_{ab\bar 4}
+C^{(4)}_{b4\bar a}- C^{(4)\dagger}_{ b4\bar a}
+C^{(4)}_{a4\bar b} -C^{(4)\dagger}_{a4\bar b}\big),
\nonumber \\
&\tilde C_{4a+4b+4} = -3 \big(
C^{(1)}_{ab4}+C^{(1)\dagger}_{ab4}
+C^{(4)}_{ab\bar 4}+ C^{(4)\dagger}_{ab\bar 4}
-C^{(4)}_{b4\bar a}- C^{(4)\dagger}_{ b4\bar a}
+C^{(4)}_{a4\bar b} +C^{(4)\dagger}_{a4\bar b}\big),
\nonumber \\
&\tilde C_{abc} =
C^{(1)}_{abc}+C^{(1)\dagger}_{abc}
+C^{(4)}_{ab\bar c}+ C^{(4)\dagger}_{ab\bar c}
-C^{(4)}_{ac\bar b}- C^{(4)\dagger}_{ ac\bar b}
+C^{(4)}_{bc\bar a} +C^{(4)\dagger}_{bc\bar a},
\nonumber \\
&\tilde C_{abc+4} =i\big(
C^{(1)}_{abc}- C^{(1)\dagger}_{abc}
-C^{(4)}_{ab\bar c}+ C^{(4)\dagger}_{ab\bar c}
-C^{(4)}_{ac\bar b}+ C^{(4)\dagger}_{ ac\bar b}
+C^{(4)}_{bc\bar a} -C^{(4)\dagger}_{bc\bar a}\big),
\nonumber \\
&\tilde C_{ab+4c+4} =-\big(
C^{(1)}_{abc}+ C^{(1)\dagger}_{abc}
-C^{(4)}_{ab\bar c}- C^{(4)\dagger}_{ab\bar c}
+C^{(4)}_{ac\bar b}+ C^{(4)\dagger}_{ ac\bar b}
+C^{(4)}_{bc\bar a} +C^{(4)\dagger}_{bc\bar a}\big),
\nonumber \\
&\tilde C_{a+4b+4c+4} =-i\big(
C^{(1)}_{abc}- C^{(1)\dagger}_{abc}
-C^{(4)}_{ab\bar c}+ C^{(4)\dagger}_{ab\bar c}
+C^{(4)}_{ac\bar b}- C^{(4)\dagger}_{ ac\bar b}
-C^{(4)}_{bc\bar a} +C^{(4)\dagger}_{bc\bar a}\big).
\end{align}

Inserting \eqref{C3-1}, \eqref{C2-4}, \eqref{C3-2}, and  \eqref{C3-3} into
\eqref{act3}, we obtain
\begin{align}\label{C3-col}
S_{\tilde C}^{(3)}=\mu_2&\int d^3x\, \frac{1}{3!}
\epsilon^{\mu\nu\rho}\Big(\tilde C_{\mu\nu\rho}+3\lambda(\tilde
C_{\mu\nu i}\tilde D_\rho\tilde X^i+v\tilde
B_{\mu\nu}A_\rho^-)+3\lambda^2(\tilde C_{\mu ij}\tilde D_\nu \tilde
X^i\tilde D_\rho\tilde X^j\nonumber\\
&+v\tilde B_{\mu i}
\langle\hspace{-0.7mm}\langle A_\nu^-\tilde D_\rho
\tilde X^i \rangle\hspace{-0.7mm}\rangle)
+\lambda^3(\tilde C_{ijk}\tilde D_\mu\tilde X^{i} \tilde
D_\nu\tilde X^{j}\tilde D_\rho\tilde X^{k}+v\tilde
B_{ij}\langle\hspace{-0.7mm}\langle A_\mu^-\tilde D_\nu\tilde X^{i}\tilde D_\rho\tilde
X^{j}\rangle\hspace{-0.7mm}\rangle)\Big).
\end{align}
On the other hand, after compactification the bosonic part of
the original U($N$)$\times$U($N$) ABJM action is given by
\begin{align}\label{oriABJM}
\int d^3x \, {\rm Tr} \Big(-\tilde D_\mu \tilde X^i \tilde
D^\mu \tilde X^i-v^2A_\mu^-A^{-\mu}+\frac k{2\pi}\
\epsilon^{\mu\nu\rho}A^-_\mu\tilde F_{\nu\rho} - V_{{\rm bos}}\Big)
+{\cal O}\big(\frac 1v\big),
\end{align}
where ${\tilde F}_{\mu\nu}=\partial_\mu A^+_\nu-\partial_\nu A^+_\mu
+ i [A^+_\mu,\, A^+_\nu]$ and $V_{{\rm bos}}$ is the reduction of
the sextic potential of ABJM theory.
As we shall discuss in the next subsection, in the reduction of
the 6-form coupling $A_\mu^-$--dependent terms are absent.
Therefore, solving the equation of motion for $A_\mu^-$
from \eqref{C3-col} and \eqref{oriABJM} gives
\begin{align}\label{C3Au}
A^{-}_\mu =\frac k{4\pi v^2}\epsilon_\mu\!\!~^{\nu\rho}\Big( {\tilde
F}_{\nu\rho}+{\mu_2 v\lambda}\frac{2\pi}k P[\tilde
B_{\nu\rho}]\Big)=\frac{1}{2g_{{\rm YM}} v}\,\epsilon_\mu^{~\nu\rho}
\Big({ \tilde F}_{\nu\rho}+\frac{1}{\tilde\lambda} P[\tilde
B_{\nu\rho}]\Big),
\end{align}
where
\begin{align}
P[\tilde B_{\mu\nu}]=\frac12\Big(\tilde B_{\mu\nu}+\lambda
\langle\hspace{-0.7mm}\langle \tilde B_{\mu i} \tilde D_\nu \tilde X^i
\rangle\hspace{-0.7mm}\rangle
+\frac{\lambda^2}3\langle\hspace{-0.7mm}\langle \tilde B_{ij}
\tilde D_\nu \tilde X^i\tilde D_\rho \tilde X^j
\rangle\hspace{-0.7mm}\rangle\Big).
\end{align}
Finally, inserting \eqref{C3Au} into the equations (\ref{C3-col})
and (\ref{oriABJM}) and rescaling the scalar field as
$\tilde X^i\to \frac {\tilde X^i}{g_{\rm YM}}$, we obtain
the following 10-dimensional action,
\begin{align}\label{C3-fin}
\int d^3x&\Bigg\{\,\frac1{g^2_{{\rm YM}}}\Big[-\tilde D_\mu \tilde X^i \tilde D^\mu
\tilde X^i-\frac12\Big({ \tilde F}_{\mu\nu}+\frac{1}{\tilde\lambda}
P[\tilde B_{\mu\nu}]\Big)^2 + \frac18[\tilde X^i,\,\tilde X^j]^2\Big]
\nonumber\\
&+\frac{\mu_2}{3!}
\epsilon^{\mu\nu\rho}\big(\tilde C_{\mu\nu\rho}+3\tilde
\lambda\tilde C_{\mu\nu i}\tilde D_\rho\tilde X^i+3\tilde
\lambda^2\tilde C_{\mu ij}\tilde D_\nu \tilde X^i\tilde D_\rho\tilde
X^j+\tilde \lambda^3\tilde C_{ijk}\tilde D_\mu\tilde X^{i}
\tilde D_\nu\tilde X^{j}\tilde D_\rho\tilde
X^{k}\big) \Bigg\}.
\end{align}
Here we notice that in addition to the natural couplings
of the D2-brane to the R-R 3-form fields in type IIA string theory,
our WZ-type coupling \eqref{act3} for the 3-form fields also
produces the coupling between $\tilde{F}_{\mu\nu}$ and $\tilde{B}_{\mu\nu}$
in the linearized non-Abelian DBI action for D2-branes.

\subsubsection{6-form fields}

In this subsection we discuss the coupling to the 6-form fields in the action
\eqref{act6}. Later we will show that the dimensionless parameter
$\tau$ we introduced in \eqref{act6} is proportional to
$1/k$. Therefore, in the doubling scaling limit $v, k\to\infty$ with fixed $v/k$,
only the terms of order $v$ or higher  are
nonvanishing in the 6-form field action.
To the leading order of $v$, the first term in \eqref{act6} is reduced to
\begin{align}\label{C6-1}
\epsilon^{\mu\nu\rho}\{{\rm Tr}\} (C_{\mu\nu\rho AB\bar C }\,\beta^{AB}_{~C})
+{\rm (c.c.)} =vi\epsilon^{\mu\nu\rho} {\rm Tr}(\tilde C_{\mu\nu\rho
ij}[\tilde X^i,\tilde X^j]),
\end{align}
where we identified the R-R 5-form fields $\tilde C_{\mu\nu\rho ij}$ as
\begin{align}
&\tilde C_{\mu\nu\rho a4}=-\frac i2(C_{\mu\nu\rho a4\bar4}
-C_{\mu\nu\rho a4 \bar4}^\dagger),\quad \tilde C_{\mu\nu\rho 4a+4}
=-\frac 12(C_{\mu\nu\rho a4\bar4 }+C_{\mu\nu\rho a 4\bar 4}^\dagger),
\nonumber\\
&\tilde C_{\mu\nu\rho ab}=-\frac i4\big(C_{\mu\nu\rho a 4\bar b }
-C_{\mu\nu\rho a 4\bar b}^\dagger -C_{\mu\nu\rho b 4\bar a}
+C_{\mu\nu\rho b 4\bar a}^\dagger +C_{\mu\nu\rho ab\bar4}
-C_{\mu\nu\rho ab\bar4}^\dagger \big),
\nonumber\\
&\tilde C_{\mu\nu\rho ab+4}=-\frac 14\big(C_{\mu\nu\rho a4\bar b}
+C_{\mu\nu\rho a4\bar b}^\dagger +C_{\mu\nu\rho b 4\bar a}
+C_{\mu\nu\rho b 4\bar a}^\dagger - C_{\mu\nu\rho ab\bar4} -
C_{\mu\nu\rho ab\bar4}^\dagger \big),
\nonumber\\
&\tilde C_{\mu\nu\rho a+4b+4}= -\frac i4\big(C_{\mu\nu\rho a4\bar
b} -C_{\mu\nu\rho a4\bar b}^\dagger -C_{\mu\nu\rho b 4\bar a}
+C_{\mu\nu\rho b 4\bar a}^\dagger -C_{\mu\nu\rho ab\bar4} +
C_{\mu\nu\rho ab\bar4}^\dagger \big).
\end{align}
For the same reason as in the paragraph below \eqref{C34},
here also we have dropped the terms involving $C_{\mu\nu\rho44\bar 4}$.
We note the fact that $C_{\mu\nu\rho AB\bar C}$ are  antisymmetric among the unbar
indices $C_{\mu\nu\rho AB\bar C} =C_{\mu\nu\rho [AB]\bar C}$ is enough
to antisymmetrize $C_{\mu\nu\rho ij}$.

Next we consider the $\lambda$-order terms in (\ref{act6}).
After the breaking of the gauge symmetry these terms can be written as
\begin{align}\label{C6-3}
\epsilon^{\mu\nu\rho}\{{\rm Tr}\}\big(&C_{\mu\nu ABC\bar D}
D_\rho Y^A \beta^{BC}_{~D} +C_{\mu\nu AB\bar C\bar D}
D_\rho Y^\dagger_C \beta^{AB}_{~D}\big) +{\rm (c.c.)}
\nonumber\\
=\epsilon^{\mu\nu\rho}{\rm Tr}\big(&C^{(1)}_{\mu\nu ABC\bar D} D_\rho Y^A
\beta^{BC}_{~D}+C^{(2)}_{\mu\nu ABC\bar D} \beta^{AB}_{~D}
D_\rho Y^C +C^{(3)}_{\mu\nu AB\bar C\bar D}D_\rho
Y^\dagger_C \beta^{AB}_{~D}\nonumber\\
&+C^{(4)}_{\mu\nu AB\bar C\bar D} \beta^{AB}_{~C}D_\rho
Y^\dagger_D\big)+{\rm (c.c.)}.
\end{align}
Keeping only the terms of order $v$ or higher and dropping the terms
involving $C_{\mu\nu 444\bar 4}$, $C_{\mu\nu 44\bar 4\bar 4}$,
$C_{\mu\nu a44\bar 4}$, $C_{\mu\nu a4\bar4\bar 4},...$
according to the logic in \eqref{C6-1}, we have
\begin{align}\label{C6-2}
\epsilon^{\mu\nu\rho}\{{\rm Tr}\}\big(&C_{\mu\nu ABC\bar D}
D_\rho Y^A \beta^{BC}_{~D}+C_{\mu\nu AB\bar C\bar D} D_\rho Y^\dagger_C
\beta^{AB}_{~D}\big)+{\rm (c.c.)}
\nonumber\\
=\epsilon^{\mu\nu\rho}{\rm Tr}\big(&2C^{(1)}_{\mu\nu ba4\bar 4} D_\rho Y^b
\beta^{a4}_{~4}+2C^{(1)}_{\mu\nu 4a4\bar b} D_\rho Y^4 \beta^{a4}_{~b}
+2C^{(1)}_{\mu\nu ca4\bar b} D_\rho Y^c \beta^{a4}_{~b}
\nonumber\\
&+C^{(1)}_{\mu\nu 4ab\bar 4} D_\rho Y^4 \beta^{ab}_{~4}
+C^{(1)}_{\mu\nu cab\bar 4} D_\rho Y^c \beta^{ab}_{~4}
+2C^{(2)}_{\mu\nu a4b\bar 4} \beta^{a4}_{~4}D_\rho Y^b
\nonumber\\
&+2C^{(2)}_{\mu\nu a44\bar b} \beta^{a4}_{~b}D_\rho
Y^4+2C^{(2)}_{\mu\nu a4c\bar b} \beta^{a4}_{~b}D_\rho
Y^c+C^{(2)}_{\mu\nu ab4\bar 4} \beta^{ab}_{~4}D_\rho
Y^4\nonumber\\
&+C^{(2)}_{\mu\nu abc\bar 4} \beta^{ab}_{~4}D_\rho Y^c
+2C^{(3)}_{\mu\nu a4\bar b\bar 4}D_\rho Y^\dagger_b \beta^{a4}_{~4}
+2C^{(3)}_{\mu\nu a4\bar 4\bar b}D_\rho Y^\dagger_4 \beta^{a4}_{~b}
\nonumber\\
&+2C^{(3)}_{\mu\nu a4\bar c\bar b}D_\rho Y^\dagger_c \beta^{a4}_{~b}
+C^{(3)}_{\mu\nu ab\bar 4\bar 4}D_\rho Y^\dagger_4 \beta^{ab}_{~4}
+C^{(3)}_{\mu\nu ab\bar c\bar 4}D_\rho Y^\dagger_c \beta^{ab}_{~4}
\nonumber\\
&+2C^{(4)}_{\mu\nu a4\bar 4\bar b}\beta^{a4}_{~4} D_\rho Y^\dagger_b
+2C^{(4)}_{\mu\nu a4\bar b\bar 4} \beta^{a4}_{~b}D_\rho Y^\dagger_4
+2C^{(4)}_{\mu\nu a4\bar b\bar c} \beta^{a4}_{~b}D_\rho Y^\dagger_c
\nonumber\\
&+C^{(4)}_{\mu\nu ab\bar 4\bar 4} \beta^{ab}_{~4}D_\rho Y^\dagger_4
+C^{(4)}_{\mu\nu ab\bar 4\bar c} \beta^{ab}_{~4}D_\rho Y^\dagger_c\big)
+{\rm (c.c.)}.
\end{align}
These terms should produce the coupling of R-R 5-form fields to D2-branes,
however substitution of \eqref{covder} and \eqref{redbeta} into \eqref{C6-2}
simultaneously produces unwanted terms.
To eliminate the unwanted terms, we will follow the same diagnoses as in the
case of 3-form coupling. Specifically, we require that the R-R 5-form field
is totally antisymmetric and the coefficients of the unwanted terms set
to be zero. We also require that the products of
$\tilde D_\mu \tilde X^i$ and $[\tilde X^j,\tilde X^k]$ are symmetrized.
For instance, the coefficients of
$\tilde D_\rho \tilde X^i[\tilde X^j,\tilde X^k]$ and $[\tilde X^j,\tilde X^k]
\tilde D_\rho \tilde X^i$ terms are identified.
In general, the 6-form fields $C^{(1)}, C^{(2)}, C^{(3)}$, and
$C^{(4)}$, introduced in \eqref{C6-3}, are not necessarily antisymmetric in
the global indices.
However, in order to match the degrees of freedom, we can consistently
impose antisymmetric property on some of 6-form fields.
In particular, we assign the antisymmetric property in the global indices
$a,b,c$ to the following 6-form fields,
\begin{align}\label{C6-4}
&C^{(1)}_{\mu\nu b[a4]\bar4},~~~ C^{(1)}_{\mu\nu
4ab\bar4},~~~C^{(1)}_{\mu\nu abc\bar4},~~~C^{(2)}_{\mu\nu
[a4]b\bar4},~~~C^{(2)}_{\mu\nu ab4\bar4},~~~C^{(2)}_{\mu\nu
abc\bar4}\nonumber\\
&C^{(3)}_{\mu\nu [a4]\bar b\bar4},~~~~C^{(3)}_{\mu\nu ab\bar
4\bar4},~~~~C^{(3)}_{\mu\nu ab\bar c\bar4},~~~~C^{(4)}_{\mu\nu
[a4]\bar 4\bar b},~~~~C^{(4)}_{\mu\nu ab\bar 4\bar
4},~~~~C^{(4)}_{\mu\nu ab\bar 4\bar c},
\end{align}
where $[a4]$ denotes antisymmetric in
interchange of $a$ and 4.
Taking these constraints into account, we obtain the following
conditions for the remaining 6-form fields:
\begin{align}\label{C6-5}
&C^{(1)}_{\mu\nu4ab\bar 4}=C^{(2)}_{\mu\nu ab4\bar 4}
=C^{(3)}_{\mu\nu ab\bar 4\bar 4}=C^{(4)}_{\mu\nu ab\bar 4\bar 4},
\quad
C^{(3)}_{\mu\nu a4\bar b \bar 4}= C^{(4)}_{\mu\nu a4\bar 4 \bar b}
= -C^{(4)\dagger}_{\mu\nu a4\bar 4 \bar b},
\nonumber \\
&C^{(1)}_{\mu\nu ab4\bar 4}= C^{(2)}_{\mu\nu a4b \bar 4}
= \frac34 C^{(4)}_{\mu\nu ab\bar 4 \bar 4} +
\frac14 C^{(4)\dagger}_{\mu\nu ab\bar 4 \bar 4},\quad
C^{(2)}_{\mu\nu a44 \bar b} + C^{(3)\dagger}_{\mu\nu b4\bar 4 \bar a}
= 2C^{(4)\dagger}_{\mu\nu a4\bar 4 \bar b},
\nonumber \\
&C^{(1)}_{\mu\nu 4a 4  \bar b}= -2 C^{(4)}_{\mu\nu a4\bar 4 \bar b}-
C^{(4)\dagger}_{\mu\nu b4\bar a \bar 4},
\quad C^{(1)}_{\mu\nu abc \bar 4}=C^{(2)}_{\mu\nu abc \bar 4},\quad
C^{(3)}_{\mu\nu ab\bar c \bar 4}=C^{(4)}_{\mu\nu ab\bar 4 \bar c},
\nonumber \\
&C^{(1)}_{\mu\nu ab4\bar c} +
C^{(4)\dagger}_{\mu\nu c4\bar b\bar a}
=- C^{(2)}_{\mu\nu a4b\bar c} - C^{(3)\dagger}_{\mu\nu c4\bar b \bar a}
= \frac12 C^{(2)\dagger}_{\mu\nu abc\bar 4} +
   \frac32 C^{(4)}_{\mu\nu ab\bar 4\bar c}.
\end{align}
Using the conditions in \eqref{C6-4} and \eqref{C6-5}, we get
\begin{align}\label{C6-sec}
\epsilon^{\mu\nu\rho}\{{\rm Tr}\}&\big(C_{\mu\nu ABC\bar D} D_\rho Y^A
\beta^{BC}_{~D}+C_{\mu\nu AB\bar C\bar D}D_\rho Y^\dagger_C
\beta^{AB}_{~D}\big)+{\rm (c.c.)}
\nonumber\\
&=vi\epsilon^{\mu\nu\rho}{\rm Tr}\big(\tilde C_{\mu\nu
ijk}\langle\hspace{-0.7mm}\langle[\tilde X^i,\tilde X^j] \tilde
D_\rho\tilde X^k\rangle\hspace{-0.7mm}\rangle\big),
\end{align}
where we have identified the R-R 5-form fields as
\begin{align}
&\tilde C_{\mu\nu ab4} =-i\big(C^{(4)}_{\mu\nu ab\bar 4\bar 4}
+ 4 C^{(4)\dagger}_{\mu\nu a4\bar 4\bar b}
- C^{(4)\dagger}_{\mu\nu ab\bar 4\bar 4}\big),\quad
\tilde C_{\mu\nu a4b+4} =-2\big(C^{(4)}_{\mu\nu ab\bar 4\bar 4}
+ C^{(4)\dagger}_{\mu\nu ab\bar 4\bar 4}\big),
\nonumber \\
&\tilde C_{\mu\nu 4a+4b+4} =i\big(
C^{(4)}_{\mu\nu ab\bar 4\bar 4} - 4 C^{(4)\dagger}_{\mu\nu a4\bar 4\bar b}
- C^{(4)\dagger}_{\mu\nu ab\bar 4\bar 4}\big),\quad
\tilde C_{\mu\nu abc} =-2i\big(C^{(4)}_{\mu\nu ab\bar 4\bar c}
- C^{(4)\dagger}_{\mu\nu ab\bar 4\bar c}\big),
\nonumber\\
& \tilde C_{\mu\nu abc+4} = C^{(2)}_{\mu\nu abc\bar 4}
+ C^{(2)\dagger}_{\mu\nu abc\bar 4}
+ C^{(4)}_{\mu\nu ab\bar 4\bar c}
+ C^{(4)\dagger}_{\mu\nu ab\bar 4\bar c},
\nonumber\\
& \tilde C_{\mu\nu ab+4c+4} =i\big(
C^{(2)}_{\mu\nu abc\bar 4} -
   C^{(2)\dagger}_{\mu\nu abc\bar 4} -
   C^{(4)}_{\mu\nu ab\bar 4\bar c} +
   C^{(4)\dagger}_{\mu\nu ab\bar 4\bar c}\big),
\nonumber\\
& \tilde C_{\mu\nu a+4b+4c+4} =
2\big( C^{(4)}_{\mu\nu ab\bar 4\bar c} +
   C^{(4)\dagger}_{\mu\nu ab\bar 4\bar c}\big).
\end{align}

The next step is to consider the reduction of the $\lambda^2$-order couplings in
\eqref{act6}. After the breakdown of the gauge symmetry, keeping only
terms which are leading order in $v$, we can write these couplings as
\begin{align}\label{C6second}
&\epsilon^{\mu\nu\rho}\{{\rm Tr}\}\big(C_{\mu ABCD\bar E} D_\nu Y^A D_\rho Y^B
\beta^{CD}_{~E} +C_{\mu ABC\bar D\bar E} D_\nu Y^A D_\rho
Y_D^\dagger \beta^{BC}_{~E}
\nonumber \\
&\hskip 1.8cm +C_{\mu AB\bar C\bar D\bar E} D_\nu Y_C^\dagger D_\rho
Y_D^\dagger \beta^{AB}_{~E} \big)+(c.c.)
\nonumber \\
&= \epsilon^{\mu\nu\rho}{\rm Tr}\big(C^{(1)}_{\mu ABCD\bar E}
D_\nu Y^A\beta^{BC}_{~E} D_\rho Y^D
+C^{(2)}_{\mu ABC\bar D\bar E}D_\nu Y^A\beta^{BC}_{~D} D_\rho Y_E^\dagger
\nonumber \\
&\hskip 1.8 cm
+C^{(3)}_{\mu ABC\bar D\bar E}D_\nu Y_D^\dagger\beta^{AB}_{~E} D_\rho Y^C
+C^{(4)}_{\mu AB\bar C \bar D \bar E}D_\nu Y_C^\dagger\beta^{AB}_{~D}
D_\rho Y_E^\dagger \big)+ ({\rm c.c.})
\nonumber \\
&\hskip 1.8cm + ({\rm permutations}),
\end{align}
where `permutations' represents the terms which can be obtained
by interchanging $\beta^{AB}_{~C}$ and covariant derivatives.
Since the terms which are obtained by the permutations do not mix
with $DY\beta DY$ terms, we will explicitly calculate only $DY\beta DY$ terms
and then symmetrize the final results.

As we did previously, we keep only terms of
order $v$ or higher and also drop terms involving 6-form fields such as
$C_{\mu ab444}$, $C_{\mu a4444}$, $C_{\mu 44444}$. Then we obtain
\begin{align}\label{secC6}
&\epsilon^{\mu\nu\rho}{\rm Tr}\big(C^{(1)}_{\mu ABCD\bar E}
D_\nu Y^A\beta^{BC}_{~E} D_\rho Y^D
+C^{(2)}_{\mu ABC\bar D\bar E}D_\nu Y^A\beta^{BC}_{~D} D_\rho Y_E^\dagger
\nonumber \\
&\hskip 0.8 cm
+C^{(3)}_{\mu ABC\bar D\bar E}D_\nu Y_D^\dagger\beta^{AB}_{~E} D_\rho Y^C
+C^{(4)}_{\mu AB\bar C\bar D \bar E}D_\nu Y_C^\dagger\beta^{AB}_{~D}
D_\rho Y_E^\dagger \big) + {\rm (c.c.)}
\nonumber \\
=&\epsilon^{\mu\nu\rho}{\rm Tr}
\big( 2  C^{(1)}_{\mu ba4c\bar 4 }  D_\nu Y^b  \beta^{a4}_{~4}  D_\rho Y^c
+ 2  C^{(1)}_{\mu ca44 \bar b}  D_\nu Y^c  \beta^{a4}_{~b}  D_\rho Y^4
+ 2  C^{(1)}_{\mu 4a4c\bar b}  D_\nu Y^4  \beta^{a4}_{~b}  D_\rho Y^c
\nonumber \\
&\hskip 0.8cm + 2  C^{(1)}_{\mu ca4d\bar b}  D_\nu Y^c  \beta^{a4}_{~b}  D_\rho Y^d
+ C^{(1)}_{\mu cab4 \bar 4}  D_\nu Y^c  \beta^{ab}_{~4}  D_\rho Y^4
+ C^{(1)}_{\mu 4abc\bar 4}  D_\nu Y^4  \beta^{ab}_{~4}  D_\rho Y^c
\nonumber \\
&\hskip 0.8cm + C^{(1)}_{\mu 4abc\bar 4}  D_\nu Y^c  \beta^{ab}_{~4}  D_\rho Y^d
+ 2  C^{(2)}_{\mu ba4\bar 4\bar c}  D_\nu Y^b  \beta^{a4}_{~4}  D_\rho Y_c^\dagger
 + 2  C^{(2)}_{\mu ca4\bar b\bar 4}  D_\nu Y^c  \beta^{a4}_{~b}  D_\rho Y_d^\dagger
\nonumber \\
&\hskip 0.8cm  + 2  C^{(2)}_{\mu 4a4\bar b\bar c}  D_\nu Y^4  \beta^{a4}_{~b}
D_\rho Y_c^\dagger
 + 2  C^{(2)}_{\mu ca4\bar b\bar d}  D_\nu Y^c  \beta^{a4}_{~b}  D_\rho Y_d^\dagger
 + C^{(2)}_{\mu cab\bar 4\bar 4}  D_\nu Y^c  \beta^{ab}_{~4}  D_\rho Y_d^\dagger
\nonumber \\
&\hskip 0.8cm
+ C^{(2)}_{\mu 4ab\bar 4\bar c}  D_\nu Y^4  \beta^{ab}_{~4}  D_\rho Y_c^\dagger
+ C^{(2)}_{\mu cab\bar 4\bar d}  D_\nu Y^c \beta^{ab}_{~4}  D_\rho Y_d^\dagger
+ 2  C^{(3)}_{\mu a4c\bar b\bar 4}  D_\nu Y_b^\dagger  \beta^{a4}_{~4}  D_\rho Y^c
\nonumber \\
&\hskip 0.8cm
+ 2  C^{(3)}_{\mu a44\bar c\bar b }  D_\nu Y_c^\dagger  \beta^{a4}_{~b}  D_\rho Y^4
+2  C^{(3)}_{\mu a4c\bar 4\bar b}  D_\nu Y_4^\dagger  \beta^{a4}_{~b}  D_\rho Y^c
+2  C^{(3)}_{\mu a4d\bar c\bar b}  D_\nu Y_c^\dagger  \beta^{a4}_{~b}  D_\rho Y^d
\nonumber \\
&\hskip 0.8cm
+C^{(3)}_{\mu ab4\bar c\bar 4}  D_\nu Y_c^\dagger  \beta^{ab}_{~4}  D_\rho Y^4
+C^{(3)}_{\mu abc\bar 4\bar 4}  D_\nu Y_4^\dagger  \beta^{ab}_{~4}  D_\rho Y^c
+C^{(3)}_{\mu abd\bar c\bar 4}  D_\nu Y_c^\dagger  \beta^{ab}_{~4}  D_\rho Y^d
\nonumber \\
&\hskip 0.8cm
+2  C^{(4)}_{\mu a4\bar c\bar b\bar 4}
D_\nu Y_c^\dagger  \beta^{a4}_{~b} D_\rho Y_d^\dagger
+2  C^{(4)}_{\mu a4\bar b\bar 4\bar c}
D_\nu Y_b^\dagger  \beta^{a4}_{~4}  D_\rho Y_c^\dagger
+2  C^{(4)}_{\mu a4\bar 4\bar b\bar c}
D_\nu Y_4^\dagger  \beta^{a4}_{~b}  D_\rho Y_c^\dagger
\nonumber \\
&\hskip 0.8cm
+2  C^{(4)}_{\mu a4\bar c\bar b\bar d}
D_\nu Y_c^\dagger  \beta^{a4}_{~b}  D_\rho Y_d^\dagger
+C^{(4)}_{\mu ab\bar 4\bar 4\bar c}
D_\nu Y_4^\dagger  \beta^{ab}_{~4}  D_\rho Y_c^\dagger
+C^{(4)}_{\mu ab\bar c\bar 4\bar c}
D_\nu Y_c^\dagger  \beta^{ab}_{~4}  D_\rho Y_d^\dagger
\nonumber \\
&\hskip 0.8cm
+ C^{(4)}_{\mu ab\bar c\bar 4\bar d}
D_\nu Y_c^\dagger  \beta^{ab}_{~4}  D_\rho Y_d^\dagger\big)+
{\rm (c.c.)}.
\end{align}

Following the argument in paragraph after \eqref{C6-2}
we impose antisymmetric property to the global indices, $a,b,c,d$,
for the following 6-form fields,
\begin{align}
&C^{(1)}_{\mu c a b4\bar 4},\, C^{(1)}_{\mu c [a 4]b\bar 4},\,
C^{(1)}_{\mu 4 a bc\bar 4},\,
C^{(2)}_{\mu cab\bar 4 \bar 4},\,C^{(2)}_{\mu b[a4]\bar 4 \bar c},\,
C^{(2)}_{\mu 4ab\bar 4 \bar c},\,
C^{(3)}_{\mu abc\bar 4 \bar 4},\,C^{(3)}_{\mu ab4\bar c \bar 4},\,
\nonumber \\
&C^{(3)}_{\mu [a4]b\bar c \bar 4 },\,
C^{(4)}_{\mu ab\bar c \bar 4 \bar 4},\,C^{(4)}_{\mu ab\bar 4 \bar 4 \bar c},\,
C^{(4)}_{\mu [a4]\bar b \bar 4 \bar c},
C^{(1)}_{\mu c a bd\bar 4},\,C^{(2)}_{\mu cab\bar 4\bar d},\,
C^{(3)}_{\mu abd\bar c \bar 4},\, C^{(4)}_{\mu ab\bar c \bar 4 \bar d}.
\end{align}
With theses constraints the remaining 6-form fields satisfy:
\begin{align}\label{condC62}
&C^{(3)}_{\mu abc\bar 4 \bar 4 }=-C^{(2)}_{\mu c ab\bar 4 \bar 4}
=C^{(1)}_{\mu 4 a bc\bar 4}= -C^{(1)}_{\mu ca b4\bar 4},
\nonumber \\
&C^{(4)}_{\mu a4\bar b \bar 4\bar c} = C^{(3)\dagger}_{\mu a4b\bar c\bar 4}
=-C^{(2)\dagger}_{\mu b a4\bar 4\bar c} = 2C^{(1)}_{\mu ca b4\bar 4}
-3C^{(1)}_{\mu ca 4b\bar 4},
\nonumber \\
&C^{(4)}_{\mu ab\bar c\bar 4\bar 4} = -C^{(4)}_{\mu ab\bar 4 \bar 4 \bar c}
=C^{(3)}_{\mu ab4\bar c\bar 4}= -C^{(2)}_{\mu 4ab\bar 4 \bar c} =
-3C^{(1)\dagger}_{\mu c a b 4\bar 4} + 4C^{(1)\dagger}_{\mu c a 4 b\bar 4},
\nonumber \\
&C^{(2)}_{\mu 4a4\bar b \bar c} - C^{(2)\dagger}_{\mu ab4\bar c\bar 4}
=C^{(4)}_{\mu a4\bar 4 \bar b \bar c} - C^{(1)\dagger}_{\mu ab 44\bar c}
=C^{(3)}_{\mu a44\bar b \bar c}-C^{(3)\dagger}_{\mu c4b\bar 4\bar a}
\nonumber \\
&=C^{(4)}_{\mu a4\bar b\bar c\bar 4}- C^{(1)\dagger}_{\mu 4c 4 b\bar a}
=-4C^{(1)}_{\mu c a b 4\bar 4} + 6 C^{(1)}_{\mu c a  4b\bar 4},
\nonumber \\
&C^{(3)}_{\mu abd\bar c\bar 4}= C^{(3)\dagger}_{\mu abd\bar c\bar 4}
=-C^{(2)}_{\mu cab\bar 4 \bar d} = -C^{(2)\dagger}_{\mu cab\bar 4 \bar d},
\nonumber \\
&C^{(2)}_{\mu b a4\bar c \bar d}-C^{(2)\dagger}_{\mu bc4\bar a \bar d}
= -C^{(3)}_{\mu d4a\bar b \bar c } + C^{(3)\dagger}_{\mu c4b \bar a \bar d }
=\frac14 C^{(1)}_{\mu c a bd\bar 4}-\frac14 C^{(1)\dagger}_{\mu c a bd\bar 4},
\nonumber \\
&C^{(4)}_{\mu ab\bar c \bar 4 \bar d}
=-\frac16 C^{(1)}_{\mu c a bd\bar 4}-\frac16 C^{(1)\dagger}_{\mu c a bd\bar 4},
\nonumber \\
&C^{(4)}_{\mu a4\bar b \bar c \bar d} = C^{(1)\dagger}_{\mu dc 4 a\bar b}
+ 2 C^{(2)}_{\mu c ab\bar 4 \bar d}.
\end{align}
Under these conditions \eqref{secC6} is reduced to
\begin{align}\label{C6-6}
\epsilon^{\mu\nu\rho}&{\rm Tr}\big(C^{(1)}_{\mu ABCD\bar E} D_\nu
Y^A\beta^{BC}_{~E} D_\rho Y^D +C^{(2)}_{\mu ABC\bar D\bar E}D_\nu
Y^A\beta^{BC}_{~D} D_\rho Y_E^\dagger
\nonumber \\
&+C^{(3)}_{\mu ABC\bar D\bar E}D_\nu Y_D^\dagger\beta^{AB}_{~E}
D_\rho Y^C +C^{(4)}_{\mu AB\bar C\bar D \bar E}D_\nu
Y_C^\dagger\beta^{AB}_{~D} D_\rho Y_E^\dagger \big) + ({\rm c.c.})
\nonumber \\
&= vi\epsilon^{\mu\nu\rho}{\rm Tr}\big( \tilde C_{\mu ijkl}D_\nu
\tilde X^i[\tilde X^{j},\,\tilde X^{k}] D_\rho \tilde X^{l}),
\end{align}
where we identified the R-R 5-form fields as
\begin{align}\label{5form2}
&\tilde C_{\mu abc 4}=-2i\big( 3C^{(1)}_{\mu c a\bar 4 b 4} - 3
C^{(1)\dagger}_{\mu c a b 4\bar 4} - 4  C^{(1)}_{\mu c a 4 b\bar 4}
+ 4  C^{(1)\dagger}_{\mu c a 4 b\bar 4}\big),
\nonumber \\
&\tilde C_{\mu a b 4 c+4}= 2  \big(C^{(1)}_{\mu c a b 4\bar 4} +
C^{(1)\dagger}_{\mu c a b 4\bar 4} - 2 C^{(1)}_{\mu c a 4 b\bar
4} - 2 C^{(1)\dagger}_{\mu c a 4 b\bar 4}\big),
\nonumber \\
&\tilde C_{\mu a4b+4c+4}=-2i \big( C^{(1)}_{\mu c a b 4\bar 4} -
C^{(1)\dagger}_{\mu c a b 4\bar 4} - 2  C^{(1)}_{\mu c a 4 b\bar 4}
+ 2  C^{(1)\dagger}_{\mu c a 4 b\bar 4}\big),
\nonumber \\
&\tilde C_{\mu 4 a+4 b+4 c+4 }=2\big(  3 C^{(1)}_{\mu c a b 4\bar 4} +
3 C^{(1)\dagger}_{\mu c a b 4\bar 4} - 4 C^{(1)}_{\mu c a 4 b\bar
4} - 4 C^{(1)\dagger}_{\mu c a 4 b\bar 4}\big),
\nonumber \\
&\tilde C_{\mu abcd+4}=\frac{1 }{3} \big( C^{(1)}_{\mu c a b d\bar 4} +
 C^{(1)\dagger}_{\mu c a b d\bar 4} + 6 C^{(2)}_{\mu
c a b \bar 4\bar d}\big),
\nonumber \\
&\tilde C_{\mu a bc+4 d+4}=\frac{ i}{2}\big( C^{(1)}_{\mu c a b d\bar 4}
- C^{(1)\dagger}_{\mu c a b d\bar 4}\big),
\nonumber \\
&\tilde C_{\mu ab+4c+4d+4}=-\frac{1 }{3} \big( C^{(1)}_{\mu c a b d\bar
4} + C^{(1)\dagger}_{\mu c a b d\bar 4} -6
C^{(2)}_{\mu c ab\bar 4  \bar d}\big).
\end{align}

Finally, including the `permutations' terms in \eqref{C6second} and
identifying the R-R 5-form fields for all the terms related by permutation
of the anticommutator $[\tilde X^i,\,\tilde X^j]$ and the covariant
derivative $\tilde D_\mu \tilde X^k$, we obtain the following
symmetrized result,
\begin{align}\label{C6-7}
\epsilon^{\mu\nu\rho}&{\rm Tr}\big(C^{(1)}_{\mu ABCD\bar E} D_\nu
Y^A\beta^{BC}_{~E} D_\rho Y^D +C^{(2)}_{\mu ABC\bar D\bar E}D_\nu
Y^A\beta^{BC}_{~D} D_\rho Y_E^\dagger
+C^{(3)}_{\mu ABC\bar D\bar E}D_\nu Y_D^\dagger\beta^{AB}_{~E}
D_\rho Y^C
\nonumber \\
&~~~ +C^{(4)}_{\mu AB\bar C\bar D \bar E}D_\nu
Y_C^\dagger\beta^{AB}_{~D} D_\rho Y_E^\dagger \big) + ({\rm c.c.})
+ ({\rm permutations})
\nonumber \\
&=vi \epsilon^{\mu\nu\rho}{\rm Tr}\big(
\tilde C_{\mu ijkl}\langle\hspace{-0.7mm}\langle
[\tilde X^{i},\,\tilde X^{j}] D_\nu \tilde X^k D_\rho \tilde X^{l}
\rangle\hspace{-0.7mm}\rangle\big).
\end{align}

It remains to carry out the dimensional reduction of
the $\lambda^3$-order terms in \eqref{act6}. Due to a huge number of
possible terms, for this order we could not be able to carry out
the reduction procedure. However, looking
at the results for the $\lambda$ and $\lambda^2$-order terms, we can
easily expect that in the compactification limit the $\lambda^3$-order terms will produce
the following 10-dimensional WZ-type coupling,
\begin{align}\label{C6-undet}
&vi \epsilon^{\mu\nu\rho}{\rm Tr}\big(
\tilde C_{ijklm}\langle\hspace{-0.7mm}\langle
[\tilde X^{i},\,\tilde X^{j}]\tilde D_\mu \tilde X^k
\tilde D_\nu \tilde X^l D_\rho \tilde X^{m}
\rangle\hspace{-0.7mm}\rangle\big).
\end{align}
The relations between the R-R 5 form $\tilde C_{ijklm}$ and the 6-form
fields need explicit calculation.

Substituting the relations \eqref{C6-1}, \eqref{C6-sec}, \eqref{C6-7}, and \eqref{C6-undet} into
 the 6-form field coupling \eqref{act6} and choosing the dimensionless parameter
$\tau$ as $ -{\pi}/{k}$, we obtain the expected WZ-type coupling
for R-R 5-form fields in type IIA string theory,
\begin{align}
S_{\tilde C}^{(5)} =&-\frac{\mu_2\tilde\lambda}{2}\int d^3x\epsilon^{\mu\nu\rho}
\frac{1}{3!} \Big(i \tilde C_{\mu\nu\rho
ij}[\tilde X^i,\tilde X^j] +3i \tilde\lambda \tilde C_{\mu\nu
ijk}\langle\hspace{-0.7mm}\langle [\tilde X^i,\tilde X^j] \tilde
D_\rho \tilde X^k\rangle \hspace{-0.7mm}\rangle
\nonumber\\
&+3i\tilde\lambda^2\tilde C_{\mu ijkl}\langle \hspace{-0.7mm}\langle
[\tilde X^i,\tilde X^j] \tilde D_\nu \tilde X^k \tilde D_\rho \tilde
X^l\rangle\hspace{-0.7mm}\rangle +i\tilde\lambda^3\tilde
C_{ijklm}\langle \hspace{-0.7mm}\langle [\tilde X^i,\tilde X^j]
\tilde D_\mu \tilde X^k \tilde D_\nu \tilde X^l\tilde D_\rho \tilde
X^m\rangle\hspace{-0.7mm} \rangle\Big).
\end{align}

\section{Comments on Mass Deformations}
\label{sec5}

The supersymmetry preserving mass deformation of the ABJM theory
was introduced in Refs.~\cite{Hosomichi:2008jb,Gomis:2008vc}.
There are several methods to obtain the mass-deformed ABJM theory,
such as $\mathcal{N}=1$ superfield formalism~\cite{Hosomichi:2008jb},
$D$-term and $F$-term deformations in $\mathcal{N}=2$ superfield
formalism~\cite{Gomis:2008vc}. These different versions of
mass-deformed ABJM theory are equivalent since they are connected by
field redefinitions~\cite{Kim:2009ny}.
A different intriguing question is to identify possible source of
the supersymmetry preserving mass deformation.
In this section we will address this question for the quartic mass
deformation.

The quadratic and quartic terms of the mass deformation are given by
\begin{align}
S_\mu= \mu^2 \int d^3x
\, {\rm Tr}\big(Y^AY_A^\dagger\big)+
\frac{4\pi\mu}{k} \int & d^3x\, {\rm Tr}
\big(M_B^{~C}Y^A Y_A^\dagger Y^BY_C^\dagger
-M^B_{~C}Y_A^\dagger Y^AY_B^\dagger Y^C \big),
\label{massdefY}
\end{align}
where $M_A^{~B}={\rm diag}(1,1,-1,-1)$.
In our present conventions these deformations can be written as
\begin{align}\label{massdefY3}
S_\mu&=\mu^2\int d^3x
 \, {\rm Tr}\big(Y^AY_A^\dagger\big)-
\frac{2\pi\mu}{k} \int d^3x
\, {\rm Tr}\big(T_{AB\bar C\bar D}Y_D^\dagger \beta^{AB}_{~C}\big)+({\rm c.c.}),
\end{align}
where $T_{AB\bar C\bar D}= M_A^{~D}\delta^C_B - M_B^{~D}\delta_A^{~C}$.
With an appropriate choice of the 6-form field, it is possible to identify the
quartic mass deformation action in \eqref{massdefY3} with
part of our WZ-type coupling in \eqref{act6}.
To that end let us consider the first term%
\footnote{In \cite{Lambert:2009qw}, the authors considered the second and the third terms
in \eqref{act6} instead of the first term. This is just a different
choice of gauge, which results in the same constant field strength.}
in \eqref{act6},
\begin{align}\label{massC}
S_{\mu}^{(6)} = \mu_2'\int d^3x\, &\frac{1}{3!}
\epsilon^{\mu\nu\rho} {\rm Tr}\big( C_{\mu\nu\rho A
B\bar C} \beta^{AB}_{~C}\big)+({\rm c.c.}),
\end{align}
where for this term the $\{{\rm Tr}\}$ in \eqref{act6} is equivalent to
ordinary trace.
In order to identify the quartic mass deformation with \eqref{massC}
the 6-form field should take the following structure,
\begin{align}
C_{\mu\nu\rho A B\bar C} = -\frac{2\mu}{\lambda\mu_2}
\epsilon_{\mu\nu\rho} T_{AB\bar C\bar D}Y_D^\dagger,\quad
C^\dagger_{\mu\nu\rho A B\bar C} = -\frac{2\mu}{\lambda\mu_2}
\epsilon_{\mu\nu\rho} T^\dagger_{AB\bar C\bar D}Y^D.
\end{align}
Since $T_{AB\bar C\bar D}$ is real, the constant 7-form field strength corresponding
to both these 6-form fields are the same,
\begin{align}\label{F7}
F_{\mu\nu\rho A B\bar C\bar D} =F^\dagger_{\mu\nu\rho A B\bar C\bar D} =
-\frac{2\mu}{\lambda^2\mu_2} \epsilon_{\mu\nu\rho} T_{AB\bar C\bar D}.
\end{align}
In the eight-dimensional transverse space this can be viewed as turning on a constant
(anti)-self dual 4-form field strength,
\begin{align}\label{F4}
F_{ A B\bar C\bar D} = -\frac{2\mu}{\lambda^2\mu_2}T_{AB\bar C\bar D}.
\end{align}
Therefore, in order to get the supersymmetry preserving quartic mass
deformation of the ABJM theory we need to turn on the constant
4-form field strengths in \eqref{F4} in the direction transverse to
the stack of parallel multiple M2-branes.\footnote{ Turning on the
other components of the form fields will break the supersymmetry,
partially or totally.}  This is in accordance with
Refs.~\cite{Bena:2000zb,Lin:2004nb,Lambert:2009qw,Kim:2010mr}, which
state that the mass-deformed ABJM action comes from a background
(anti)self dual 4-form flux in the eight-dimensional transverse
space. The quadratic mass-deformed term in \eqref{massdefY3} is
originated from the backreaction of the background metric in the
presence of the 4-form flux~\cite{Lambert:2009qw}.

\setcounter{equation}{0}
\section{Conclusion and Discussion}\label{sec6}

In this work we discussed the coupling of multiple M2-branes to the
background 3- and 6-form fields in 11-dimensional supergravity.
We proposed a general gauge-invariant WZ-type coupling in the ABJM theory
with U($N$)$\times$U($N$) gauge group.
The fundamental building blocks of the coupling are the 3- and 6-form gauge fields,
the covariant derivative $D_\mu Y^A$, the 3-commutator $\beta^{AB}_{~C}$,
and their complex conjugates. We assumed that
the form fields depend on the transverse scalars $Y^A$ and their complex
conjugates. Therefore, the form fields carry gauge indices according to
their dependence on $Y^A$ and $Y_A^\dagger$ to guarantee gauge invariance
which allows both single and multi traces.
Since the multi-trace terms after the MP Higgsing cannot be
identified with any WZ-type couplings in type IIA string theory,
we did not include those terms from the beginning.
We tested our proposal for the WZ-type coupling through several
consistency checks.

We compared our WZ-type coupling with the known
coupling for a single M2-brane. In this
case the gauge group is U(1)$\times$U(1). We showed that our
coupling completely matches the well known effective action of a
single M2-brane coupled to 3-form fields. Then we carried out
the MP Higgsing procedure for our WZ-type coupling including the
bosonic part of the original ABJM action with U($N$)$\times$U($N$)
gauge symmetry. We obtained the symmetrized WZ-type coupling for the
R-R 3- and 5-form fields with the gauge field strength turned
off~\cite{Myers:1999ps}.  The MP Higgsing procedure alone
is not enough to reproduce all the expected WZ-type couplings in type IIA
string theory. Therefore, we imposed some constraints on the
11-dimensional form fields, in order
\begin{itemize}
\item to make the 10-dimensional form fields antisymmetric,
\item to obtain a 10-dimensional WZ-type coupling which is symmetrized
with  respect to interchange of the covariant derivatives $(\tilde
D_\mu\tilde X^i)$ and the commutators ($[\tilde X^j\,,\tilde X^k]$), and
\item to make disappear quadratic terms in the auxiliary
gauge field $A^-_\mu$ in 3-form coupling and linear terms
in $A_\mu^-$ in 6-form coupling, which are absent in WZ-type
coupling in type IIA string theory.
\end{itemize}
We have also showed that, in addition to the WZ-type coupling for
the R-R 3- and 5-form fields, the reduction procedure also produces
correctly the linearized version of DBI action for the gauge field
strength $\tilde F_{\mu\nu}$ of the dynamical gauge field
$A_{\mu}^+$ and the NS-NS 2-form field $\tilde B_{\mu\nu}$.
Finally, we tested the validity of our proposal by establishing
the relation between the WZ-type coupling for the 6-form field and
the supersymmetry preserving quartic term of the
mass-deformed action in ABJM theory with U($N$)$\times$U($N$)
gauge symmetry.

A few comments about our construction are in order.
Though the 10-dimensional WZ-type coupling
is restricted to the linear terms in form fields,
it contains an exponential form of the coupling of the NS-NS 2-form field.
Since the NS-NS 2-form field comes from compactification
of the 3-form field in M-theory, it seems natural to include
nonlinear terms which deserves further investigations.
In addition to the fundamental building
blocks of our WZ-type coupling listed above, we can add gauge field strengths
for $A_{\mu}$ and $\hat A_\mu$ and the monopole operator, which changes
bifundamental representation to antibifundamental and vice versa
for $k=1,2$~\cite{Gustavsson:2009pm}.
With these extensions we may include extra gauge-invariant terms to our
WZ-type couplings.

\section*{Acknowledgements}
The authors would like to appreciate the informative discussions
with Jaemo Park. This work was supported by the Korea Research
Foundation Grant funded by the Korean Government  with grant number
KRF-2008-313-C00170 (Y.K.), 2009-0073775 (O.K.), and 2009-0077423
(D.D.T.). This work was also supported by the National Research
Foundation of Korea (NRF) grant funded by the Korean Government
(MEST) (No. 2009-0062869) through Astrophysical Research Center for
the Structure and Evolution of the Cosmos (ARCSEC) (Y.K.), and
by Mid-career Researcher Program
through the National Research Foundation of Korea(NRF) grant
funded by the Korea government(MEST)(No. 2009-0084601) (H.N.).

\appendix

\section{Comments on Single Trace Terms}\label{appA}
In our construction of the WZ-type actions in \eqref{act3} and
\eqref{act6} the 3- and 6-form fields are arbitrary functions of the
transverse complex scalar fields except that every term in the actions
\begin{itemize}
\item  should contain equal number of
bifundamental and antibifundamental fields to guarantee gauge
invariance,
\item and should be single trace.
\end{itemize}
After the MP Higgsing procedure, this arbitrariness results in terms
which are absent in the known WZ-type action of in
type IIA string theory. In this appendix we shall show how
to overcome this problem by imposing more constraints on the
functional dependence of the form fields on the complex scalar fields.

Let us consider the following term in \eqref{act3},
\begin{align}\label{A0}
 \{{\rm Tr}\}\big(C_{\mu A\bar B}D_\nu
Y^AD_\rho Y_B^\dagger\big).
\end{align}
For clear presentation we drop the worldvolume and global
indices from now on. Imposing the single traceness constraint on
 \eqref{A0} we have
\begin{align} \label{A2}\{{\rm Tr}\}(C DY
DY^\dagger)=C^{\hat ab}_{a\hat b} (DY)^{a}_{\hat a} (D
Y^\dagger)^{\hat b}_b={\rm Tr}(HD Y^AID Y^\dagger),
\end{align}
where $C^{\hat a b}_{a\hat b}= H^{\hat a}_{\hat b}I^b_{a}$, and both $H$
and $I$ should depend on the transverse scalar fields to guarantee the
U($N$)$\times$U($N$) gauge invariance.
If we turn on the vev for the scalar fields, $Y^A\to\frac
v2\delta^{A4}+Y^A$, it breaks the U($N$)$\times$U($N$)
gauge symmetry to a U($N$) and then both the bifundamental and
antibifundamental fields follow adjoint representation.
For those adjoint fields  the indices
$a,b,\cdots$ and $\hat a,\hat b, \cdots$ are indistinguishable so
that the matrices $H$ and $I$ are expressed as
\begin{align}\label{A3}
H^a_{~b}&= h_0\delta^a_b + h_1\tilde Y^a_{~b} + h_2 (\tilde Y\tilde
Y)^a_{~b} +\cdots,
\nonumber \\
I^a_{~b}&= i_0\delta^a_b + i_1\tilde Y^a_{~b} + i_2 (\tilde Y\tilde
Y)^a_{~b} + \cdots,
\end{align}
where $h_i$ and $i_i$ are the coefficients which can depend on the
parameters $v,k,$ and $l_{\rm P}$, and $\tilde Y$ is either
$Y$ or $Y^\dagger$.  Substituting \eqref{A3} in \eqref{A2}, we have
\begin{align}
{\rm Tr}(HD Y^AID Y^\dagger)={\rm Tr}\big(C^{(1)} DYDY^\dagger +C^{(2)} DY^\dagger DY +{\cal H}
DY {\cal I} DY^\dagger\big)\label{A1},
\end{align}
where
\begin{align}
C^{(1)} &= \big(h_0 + h_1 \tilde Y + h_2\tilde Y\tilde Y+\cdots)i_0,
\nonumber \\
C^{(2)} &= \big(i_0 + i_1 \tilde Y + i_2\tilde Y\tilde Y+\cdots)
h_0,
\nonumber \\
{\cal H} &=  h_1 \tilde Y + h_2\tilde Y\tilde Y+\cdots,
\nonumber \\
{\cal I} &= i_1 \tilde Y + i_2\tilde Y\tilde Y+\cdots.
\end{align}
In our reduction procedure in section \ref{sec4} we have dropped
the last term in \eqref{A1}. This is justified if we restrict
the form fields in string theory be at most linear in
the scalar fields. In generic setup we can realize
the last term in \eqref{A1} can appear.
Therefore, in order to reproduce the known WZ-type coupling with
arbitrary dependence on the transverse scalars in type IIA
string theory, we should find an appropriate condition on the
form field couplings in M-theory to forbid the last term in
\eqref{A1} and all the other terms of this kind.

\end{document}